\documentclass[12pt]{article}
\usepackage[english,activeacute]{babel}
\usepackage{natbib}
\usepackage{comment}
\usepackage{float}
\usepackage{hyperref}
\usepackage{mathrsfs}
\usepackage{enumitem}
\usepackage[font={small,it}]{caption}
\usepackage{amsmath,amsfonts,amsthm,amssymb}
\usepackage{bm,rotating,multirow,dsfont,graphicx}
\usepackage[usenames, dvipsnames]{color}
\usepackage{url}
\usepackage{multicol}
\usepackage{multirow}
\usepackage[T1]{fontenc}
\usepackage{flafter}
\usepackage{appendix}
\usepackage{subfigure}
\usepackage{xcolor}

\hypersetup{
	colorlinks,
	linkcolor={blue!70!black},
	citecolor={blue!70!black},
	urlcolor ={blue!70!black}
}

\makeatletter
\def\hlinewd#1{%
	\noalign{\ifnum0=`}\fi\hrule \@height #1 %
	\futurelet\reserved@a\@xhline}
\makeatother

\DeclareMathOperator*{\argmin}{arg\,min}
\newcommand\simiid{\mathrel{\overset{\makebox[0pt]{\mbox{\normalfont\tiny\sffamily iid}}}{\sim}}}
\newcommand\simind{\mathrel{\overset{\makebox[0pt]{\mbox{\normalfont\tiny\sffamily ind}}}{\sim}}}
\newcommand{\pr}[1]{\textsf{Pr}\left(#1\right)}
\newcommand{\ind}[1]{\mathbf{1}\left\{ #1 \right\}}
\newcommand\floor[1]{\lfloor#1\rfloor}
\newcommand{\expec}[1]{\textsf{E}\left(#1\right)}

\newcommand{\var}[1]{\textsf{Var}\left(#1\right)}
\newcommand{\sd}[1]{\textsf{SD}\left(#1\right)}

\def\L{\mathbf{L}}
\def\D{\mathbf{D}}
\def\R{\mathbf{R}}
\def\E{\mathbf{E}}
\def\I{\mathbf{I}}
\def\Q{\mathbf{Q}}
\def\V{\mathbf{V}}

\def\Y{\mathbf{Y}}
\def\X{\mathbf{X}}
\def\UPS{\mathbf{\Upsilon}}
\def\LAM{\mathbf{\Lambda}}
\def\mv{\boldsymbol{m}}
\def\uv{\boldsymbol{u}}
\def\vv{\boldsymbol{v}}
\def\xv{\boldsymbol{x}}
\def\kap{\kappa}
\def\sig{\sigma}

\def\etav{\boldsymbol{\eta}}
\def\tev{\boldsymbol{\theta}}
\def\nuv{\boldsymbol{\nu}}
\def\zev{\boldsymbol{\zeta}}

\def\Ber{\small{\mathsf{Ber}}}

\def\Nor{\small{\mathsf{N}}}

\def\IGamd{\small{\mathsf{IG}}}

\def\zerov{\boldsymbol{0}}
\def\onev{\boldsymbol{1}}
\def\reals{\mathbb{R}}
\def\tr{\mathsf{tr}}
\def\trans{\textsf{T}}
\def\@roman#1{\romannumeral #1}

\begin{document}

\def\spacingset#1{\renewcommand{\baselinestretch}{#1}\small\normalsize}\spacingset{1}

\title{A Latent Space Model for Multilayer Network Data}

\date{}

\author{
	Juan Sosa, Universidad Nacional de Colombia, Colombia\footnote{jcsosam@unal.edu.co} \\
	Brenda Betancourt, University of Florida, United States\footnote{bbetancourt@ufl.edu}
}	 
	       
\maketitle

\begin{abstract}
In this work, we propose a Bayesian statistical model to simultaneously characterize two or more social networks defined over a common set of actors. The key feature of the model is a hierarchical prior distribution that allows us to represent the entire system jointly, achieving a compromise between dependent and independent networks. Among others things, such a specification easily allows us to visualize multilayer network data in a low-dimensional Euclidean space, generate a weighted network that reflects the consensus affinity between actors, establish a measure of correlation between networks, assess cognitive judgements that subjects form about the relationships among actors, and perform clustering tasks at different social instances. Our model's capabilities are illustrated using several real-world data sets, taking into account different types of actors, sizes, and relations.
\end{abstract}


\spacingset{1.1} 

\section{Introduction}

The study of information that emerges from the interconnectedness among autonomous elements in a system (and the elements themselves) is extremely important in the understanding of many phenomena. Structures formed by these elements (individuals or actors) and their interactions (ties or connections), commonly known as networks, are popular in many research areas such as finance (studying alliances and conflicts among countries as part of the global economy), social science (studying interpersonal social relationships and social schemes of collaboration such as legislative cosponsorship networks), biology (studying arrangements of interacting genes, proteins or organisms), epidemiology (studying the spread of a infectious disease), and computer science (studying the Internet, the World Wide Web, and also communication networks), just to mention a few examples, primarily because interactions typically arise under several contexts or points of view.

Relational structures consisting of $J$ types of interactions (layers or views) established over a common set of $I$ actors are regularly referred to as either multilayer or multiview network data, which results in a sequence of $J$ adjacency matrices $\Y_1,\ldots,\Y_J$, with $\Y_{j} = [y_{i,i',j}]_{i,i'=1,\ldots,I, i\neq i'}$ for $j=1,\ldots,J$, each having structural zeros along the main diagonal (note that $y_{i,i',j}\equiv y_{i',i,j}$ for undirected relations). This type of data is very frequent nowadays. For instance, the interactions that employees have with others according to their roles in work are not necessarily the same as the interpersonal relationships they build among them; however, the corresponding social structures defined by these two types of relationships may have some characteristics in common. Thus, given the richness of information provided in $\Y=\{ \Y_j \}$, our main goal consists of modeling dependencies both within and between layers in order to formally test features about the social dynamics in the system.

A very popular statistical model in the literature for a single network is the latent position model given in \cite{hoff-2002}. According to this model, interaction probabilities marginally depend on how close or far apart actors are on a latent ``social space'' (a $K$-dimensional vector space, typically $\reals^K$, in which each individual occupies a fixed position). This formulation is appealing because latent structures based on distances naturally induce transitivity and homophily, which are typical features found in many social networks. Other meaningful advances in latent space models for networks can be found in \citet{nowicki-2001}, \citet{schweinberger-2003}, \citet{hoff-2005,hoff-2008, hoff-2009}, \citet{handcock-2007}, \cite{linkletter-2007}, \citet{krivitsky-2008}, \citet{krivitsky-2009}, and \citet{li-2011}.

Here, we extend Hoff's latent position model in order to describe the generative process of cross-sectional multilayer network data. The key feature of our model is a hierarchical prior distribution that allows us to characterize the entire system jointly. Such a prior specification is very convenient for multiple reasons. First, the model provides a direct description of actors' roles within and across networks at global and specific levels. Second, it provides the tools for representing several network features effortlessly at any instance. Finally, the proposed framework accounts for dependence structures between layers which is key to perform formal tests about actor and network characteristics.

Perhaps the closest in spirit to our modeling strategy is the work given in \cite{gollini2016joint} and \cite{salter-2017}. Unlike our approach, \cite{gollini2016joint} introduce a latent space model assuming that the interaction probabilities in each network view
is explained by a unique latent variable. Later, \cite{salter-2017} consider the same assumption but in the context of a multivariate Bernoulli likelihood, which leads to a clear estimate of interview dependence. Our proposal builds on the latent configuration of these models, by considering a full hierarchical prior specification that provides a parsimonious characterization of actors from many perspectives.

Aside from the previous work, other alternatives for studying multilayer network data have emerged during the last two years from the latent space modeling perspective. In brain connectomics, \cite{durante2018bayesian} present a Bayesian nonparametric approach via mixture modeling, which reduces dimensionality and efficiently incorporates network information within each mixture component by leveraging latent space representations. More recently, \cite{wang2019common} propose a method to estimate a common structure and low-dimensional individual-specific deviations from replicated networks, based on a logistic regression mapping combined with a hierarchical singular value decomposition. In turn, D'Angelo an collaborators extend latent space models in other contexts, by considering node-specific effects \citep{d2018node}, network-specific parameters and edge-specific covariates \citep{d2019latent}, and finally, a clustering structure in the framework of an infinite mixture distribution \citep{d2020model}. Other important advances from a frequentist point of view are available in \cite{zhang2020statistical}.

Additional work related to cross-sectional multilayer network data includes community detection (e.g., \citealp{han2015consistent}, \citealp{reyes2016stochastic}, \citealp{paul2016consistent}, \citealp{gao-2019}, \citealp{paez-2019}, and \citealp{paul2020spectral}),
and the perception assessment in cognitive social structures (e.g., \citealp{swartz-2015}, \citealp{sewell2019latent}, and \citealp{sosa2021cognitive}). Finally, from the dynamic point of view, there is a large variety of approaches to modeling network evolution over time (e.g., \citealp{durante2014nonparametric}, \citealp{hoff2015multilinear}, \citealp{sewell2015latent, sewell2016latent, sewell2017latent}, \citealp{gupta2018evolving}, \citealp{kim2018review}, \citealp{turnbull2020advancements}, \citealp{betancourt2020modelling}).

Our contribution has many folds. In Section \ref{sec_latent_space_models}, we present our proposal for modeling multiple layer network data, including prior elicitation. In Section \ref{sec_computation}, we discuss the topics of identifiability and model selection for our approach. Next, in Section \ref{sec_ilustrations}, we provide two illustrations using popular data sets in the literature, for which we develop formal tests involving network correlation and perceptual agreement, as well as a full analysis of the social dynamics. In Section \ref{sec_CV}, we carry out a cross-validation study on additional datasets in order to test the predictive capabilities of our proposed model. Finally, concluding remarks and directions for future work are provided in Section \ref{sec_discussion}.

\section{Latent space models}\label{sec_latent_space_models}

Since the foundational work of \cite{hoff-2002}, generalized linear mixed models became a popular alternative to model networks. In particular, consider an undirected binary network $\Y = [y_{i,i'}]$ in which the $y_{i,i'}$\,s, $i,i'=1,\ldots,I$, $i<i'$, are assumed to be conditionally independent with interaction probabilities $\vartheta_{i,i'} \equiv \pr{ y_{i,i'} = 1 \mid \zeta, \gamma_{i,i'} } = \Phi(\zeta + \gamma_{i,i'})$, where $\Phi(\cdot)$ denotes the cumulative distribution function of the standard Gaussian distribution (other link functions can be considered), $\zeta$ is a fixed effect representing the global propensity of observing an edge between actors $i$ and $i'$, and $\gamma_{i,i'}$ is an unobserved dyad-specific random effect representing any additional patterns unrelated to those captured by $\zeta$.  Following results in \citet{hoover-1982} and \citet{aldous-1985} \citep[see also][]{hoff-2008}, it can be shown that if the matrix of random effects $[\gamma_{i,i'}]$ is jointly exchangeable, there exists a symmetric function $\alpha(\cdot,\cdot)$ and a sequence of independent random variables (vectors) $\uv_1,\ldots,\uv_I$ such that $\gamma_{i,i'} = \alpha(\uv_i,\uv_{i'})$. It is mainly through $\alpha(\cdot,\cdot)$ that we are able to capture relevant features of the network. A number of potential formulations for $\alpha(\cdot,\cdot)$ have been explored in the literature to date; see \cite{minhas2019inferential} and \cite{sosa-2021} for a review.

\begin{figure}[!h]
	\centering
	\includegraphics[scale=0.77]{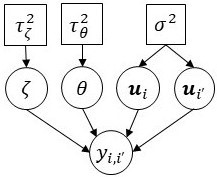}
	\caption{DAG representation of the LPM for a single network. Circles represent either random variables or random vectors, and the edges convey conditional independence. Squares represent fixed quantities (constants).}
	\label{fig_DAG_distance_model}
\end{figure}

In particular, consider the latent position model (LPM) given in \cite{hoff-2002}. This model assumes that each actor $i$ has an unknown position $\uv_i$ in a social space of latent characteristics, typically $\uv_i=(u_{i,1},\ldots,u_{i,K})\in\reals^K$, where $K$ is assumed to be known, and that the probability of an edge between two actors may decrease as the latent characteristics of the individuals become farther apart of each other. In this spirit, the latent effects can be specified as $\gamma_{i,i'}=-e^\theta\|\,\uv_i-\uv_{i'}\|\,$, where $e^\theta$ serves as a weighting factor that regulates the contribution attributed to the latent effects, and therefore, $y_{i,i'}\mid\zeta,\theta,\uv_i,\uv_{i'} \simind \Ber\left(\Phi\left(\zeta -e^\theta\|\,\uv_i-\uv_{i'}\|\,\right)\right)$. In order to perform a fully Bayesian analysis, we must specify a prior distribution on the model parameters; a standard choice that works well in practice consists in setting mutually independent prior distributions, $\zeta \sim \Nor_1(0,\tau_\zeta^2)$, $\theta \sim \Nor_1(0,\tau_\theta^2)$, and $\uv_i\simiid \Nor_K(\zerov,\sig^2\,\I)$, for constants $\tau^2_\zeta,\tau^2_\theta,\sigma^2>0$, although other similar formulations are available \citep[e.g.,][]{rastelli2019computationally}. Thus, the entire model has $IK + 2$ unknown parameters to estimate, namely, $\zeta,\theta,\uv_1,\ldots,\uv_I$, associated with the hyperparameters $\tau^2_\zeta,\tau^2_\theta$, and $\sigma^2$, which need to be picked sensibly to ensure appropriate model performance. In our experience, letting $\tau^2_\zeta = \tau^2_\theta = 3$ and $\sigma^2 = 1/9$ is a reasonable choice; however, other heuristics are possible \cite[e.g.,][]{krivitsky-2009}. Figure \ref{fig_DAG_distance_model} provides a directed acyclic graph (DAG) representation of the LPM for a single network. In the following section we present an extension of this model that is suited for multilayer networks.

\subsection{Extension to multilayer networks}\label{sec_multilayer_networks}

Here, we present our approach to simultaneously model a set of $J\geq 2$ undirected binary networks $\Y_1,\ldots,\Y_J$, defined over a common set of $I$ actors, with $\Y_j = [y_{i,i',j}]$ for $j = 1,\ldots,J$. Since each network contains relevant information about a determined aspect of the social dynamics, instead of just fitting independent LPMs to each network, the main idea behind our approach consists of borrowing information across networks by means of a hierarchical prior specification on the interaction probabilities $\vartheta_{i,i',j}$.

Our model is an unequivocal hierarchical extension of the LPM for a single network that accommodates relevant features associated with multilayer network data. Here, we still assume that observations are conditionally independent, $y_{i,i',j}\mid\vartheta_{i,i',j}\simind \Ber\left(\vartheta_{i,i',j}\right)$, and construct a hierarchical prior on the array $[\vartheta_{i,i',j}]$, by letting
\begin{equation}\label{eq_MNLPM_interaction_probabilities}
\vartheta_{i,i',j} = \Phi\left(\zeta_j -e^{\theta_j}\|\,\uv_{i,j} - \uv_{i',j}\|\,\right)\,,
\end{equation}
where the additional index $j$ makes explicit the reference to network $j$, i.e., $\zeta_j$ is the global propensity of observing an edge between actors $i$ and $i'$ in network $j$, $e^{\theta_j}$ is a weighting factor that regulates the contribution attributed to the latent effects in network $j$, and $\uv_{i,j} = (u_{i,j,1},\ldots,u_{i,j,K})$ is the latent position of actor $i$ in network $j$. In this context, note that the interpretation of the latent structure remains unchanged: if $\uv_{i,j}$ and $\uv_{i',j}$ ``move away'' from each other in the social space, then $\|\,\uv_{i,j} - \uv_{i',j}\|\,$ increases, and therefore, the probability of observing an edge between actors $i$ and $i'$ in network $j$ may decrease depending on the regularization provided by $e^{\theta_j}$.

If mutually independent prior distributions were assigned to each set of $\zeta_j$\,s, $\theta_j$\,s, and $\uv_{i,j}$\,s, then such a formulation would be equivalent to fitting independently a LPM to each network. Instead, we consider a hierarchical prior distribution that characterizes the heterogeneity of the model parameters across networks. Our approach parsimoniously places conditionally independent Gaussian priors as follows:
\begin{align}\label{eq_MLSM_hierchy_1}
\zeta_j     \mid \mu_\zeta, \tau^2_\zeta     &\simiid \Nor_1\left(\mu_\zeta, \tau^2_\zeta\right),     &
\theta_j    \mid \mu_\theta, \tau^2_\theta   &\simiid \Nor_1\left(\mu_\theta, \tau^2_\theta\right),   &
\uv_{i,j}   \mid \etav_i, \sig^2             &\simind \Nor_K\left(\etav_i, \sig^2\I\right).           
\end{align}
On the one hand, $(\mu_\zeta, \tau^2_\zeta)$ and $(\mu_\theta, \tau^2_\theta)$ parameterize the sampling distributions that describe the heterogeneity across networks in terms of fixed effects $\zeta_1,\ldots,\zeta_J$ and weighting log-factors $\theta_1,\ldots,\theta_J$, respectively. On the other hand, the mean $\etav_i = (\eta_{i,1},\ldots,\eta_{i,K})$ can be conveniently interpreted as the average ``global'' position of actor $i$ in relation to the dynamics that define social interactions in the system. Now, we can capture similarities among the observed networks and borrow information across them, mainly by placing a common prior distribution on $\etav_1,\ldots,\etav_I$. Thus, we let 
\begin{align}\label{eq_MLSM_hierchy_2}
\etav_i  \mid \nuv, \kappa^2  &\simiid \Nor_K\left(\nuv, \kappa^2\I\right), &
\sig^2                        &\sim \IGamd\left(a_\sig,b_\sig \right)\,,
\end{align}
in order to characterize between-actor mean sampling variability in a straightforward fashion. Furthermore, note that the sampling variability of the latent positions $\sigma^2$ is assumed to be constant across actors and networks. We believe this is a sensible choice because inferences on the latent positions seem to be invariant when we eliminate such an assumption. 

\begin{figure}[!h]
	\centering
	\includegraphics[scale=0.77]{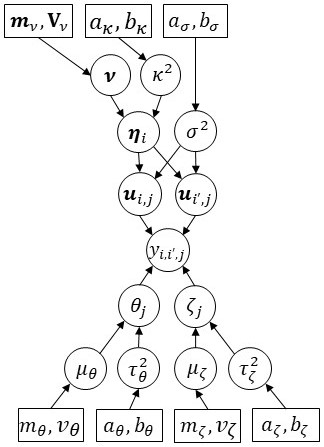}
	\caption{DAG representation of the MNLPM for multilayer network data.}
	\label{fig_DAG_multiple_distance_model}
\end{figure}

Finally, the model is completed by specifying prior distributions in a conjugate fashion on the remaining model parameters:
\begin{align}\label{eq_MLSM_hierchy_3}
\mu_\zeta       &\sim  \Nor_1\left(m_\zeta,  v^2_\zeta\right),    &
\mu_\theta      &\sim  \Nor_1\left(m_\theta, v^2_\theta\right),   &
\nuv            &\sim  \Nor_K\left(\mv_{\nuv}, \V_{\nuv} \right), \nonumber \\
\tau^2_\zeta  	&\sim  \IGamd\left(a_\zeta,b_\zeta\right),        &
\tau^2_\theta  	&\sim  \IGamd\left(a_\theta,b_\theta\right),      & 
\kappa^2        &\sim  \IGamd\left(a_\kappa, b_\kappa \right),    
\end{align}
where $a_\sig, b_\sig, a_\zeta, b_\zeta, a_\theta,b_\theta, a_\kappa, b_\kappa, m_\zeta, v_\zeta, m_\theta, v_\theta, \mv_{\nuv}, \V_{\nuv}$ are fixed hyperparameters. Therefore, the full set of model parameters is
$$
\UPS \equiv \UPS_{I,J,K} = \left(\zeta_1,\ldots,\zeta_J, \theta_1,\ldots,\theta_J, \uv_{1,1},\ldots,\uv_{I,J}, \etav_1,\ldots,\etav_I, \sigma, \mu_\zeta, \tau_\zeta, \mu_\theta, \tau_\theta, \nuv, \kappa \right)\,,
$$ 
which includes $IK(J + 1) + 2J + K + 6$ unknown quantities to estimate. Figure \ref{fig_DAG_multiple_distance_model} shows a DAG representation of our Multilayer Network Latent Position Model (MNLPM) given in \eqref{eq_MNLPM_interaction_probabilities}, \eqref{eq_MLSM_hierchy_1}, \eqref{eq_MLSM_hierchy_2}, and \eqref{eq_MLSM_hierchy_3}. Note the clear hierarchical structure in the model.

This model for multilayer network data is such that the resulting joint marginal distribution of the data is fully jointly exchangeable, which means that the joint distribution of $\{ y_{i,i',j} \}$ is the same as the distribution of $\{ y_{\pi_1(i), \pi_2(i'), \pi_3(j)} \}$ only if $\pi_1 = \pi_2$, where $\pi_1$ and $\pi_2$ are permutations of $[I]$, and $\pi_3$ is a permutation of $[J]$. Full joint exchangeability (rather than a weaker form of exchangeability) is particularly attractive in this setting because all indexes $i$, $i'$ (and potentially $j$) refer to the same set of actors \citep{sosa2021cognitive}.

\subsection{Prior elicitation}

Once again, careful elicitation of the hyperparameters is key to ensure appropriate model performance since the model is sensitive to this choice. In our experience, the following heuristic procedure produces adequate results for a wide variety of multilayer network datasets.  In the absence of prior information, we set $m_\theta = m_\zeta = 0$ and $\mv_{\nuv} = \zerov$ and $\V_{\nuv} = v_\nu^2\I$ in order to center the model, roughly speaking, around an Erd\"{o}s-R{\'e}nyi model \citep{erdos-1959}, and also ensure that the prior distributions are invariant to rotations of the latent space (see also Section \ref{sec_identifiability}). 

Now, we establish some constraints that allow us to appropriately contrast models constructed with different values of $K$. Thus, for the prior distributions on the variance parameters, we naturally let $a_\zeta = a_\theta = a_\sigma = a_\kappa = 3$, which leads to a proper prior with finite moments and a coefficient of variation equal to 1. Under this set up, it can be shown that marginally
\begin{align*}
\var{\zeta_j}   &= \frac{b_\zeta}{2} + v^2_\zeta\,, &
\var{\theta_j}  &= \frac{b_\theta}{2} + v^2_\theta\,, &
\var{u_{i,j,k}} &= \frac{b_\sig}{2} + \frac{b_\kap}{2} + v^2_\nu,
\end{align*}
each of which we split equally among all terms. First, resembling a regular LPM, we set $\var{u_{i,j,k}} = 1/9$ a priori, such that $b_\sig = b_\kap = 2/27$ and $v^2_\nu = 1/27$. Then, from a naive application of the delta method we obtain that
$$
\var{\theta_j} \doteq 2\log\left( -\frac{\Phi^{-1}(\vartheta_0)}{\expec{\|\,\uv_{i,j} - \uv_{i',j}\|\,}} \right)\,
$$
where $\vartheta_0$ is the prior probability of observing an edge between any two actors (which can be tuned to reflect prior information), and $\uv_{i,j} - \uv_{i',j}\sim \Nor_K( \zerov,\tfrac29\I )$ as long as $\var{u_{i,j,k}} = \var{\eta_{i,k}} = 1/9$. In our experiments, we set $b_\theta = \var{\theta_j}$ and $v^2_\theta = \var{\theta_j}/2$ with $\vartheta_0 = 0.1$. Finally, we set $b_\zeta = \var{\zeta_j}$ and $v^2_\zeta = \var{\zeta_j}/2$ with $\var{\zeta_j} = 4\expec{\|\,\uv_{i,j} - \uv_{i',j}\|\,}$, which allows a wide range of values of $\zeta_j$.

\begin{figure}[!t]
	\centering
	\subfigure[$K=1$] {\includegraphics[scale=0.4]{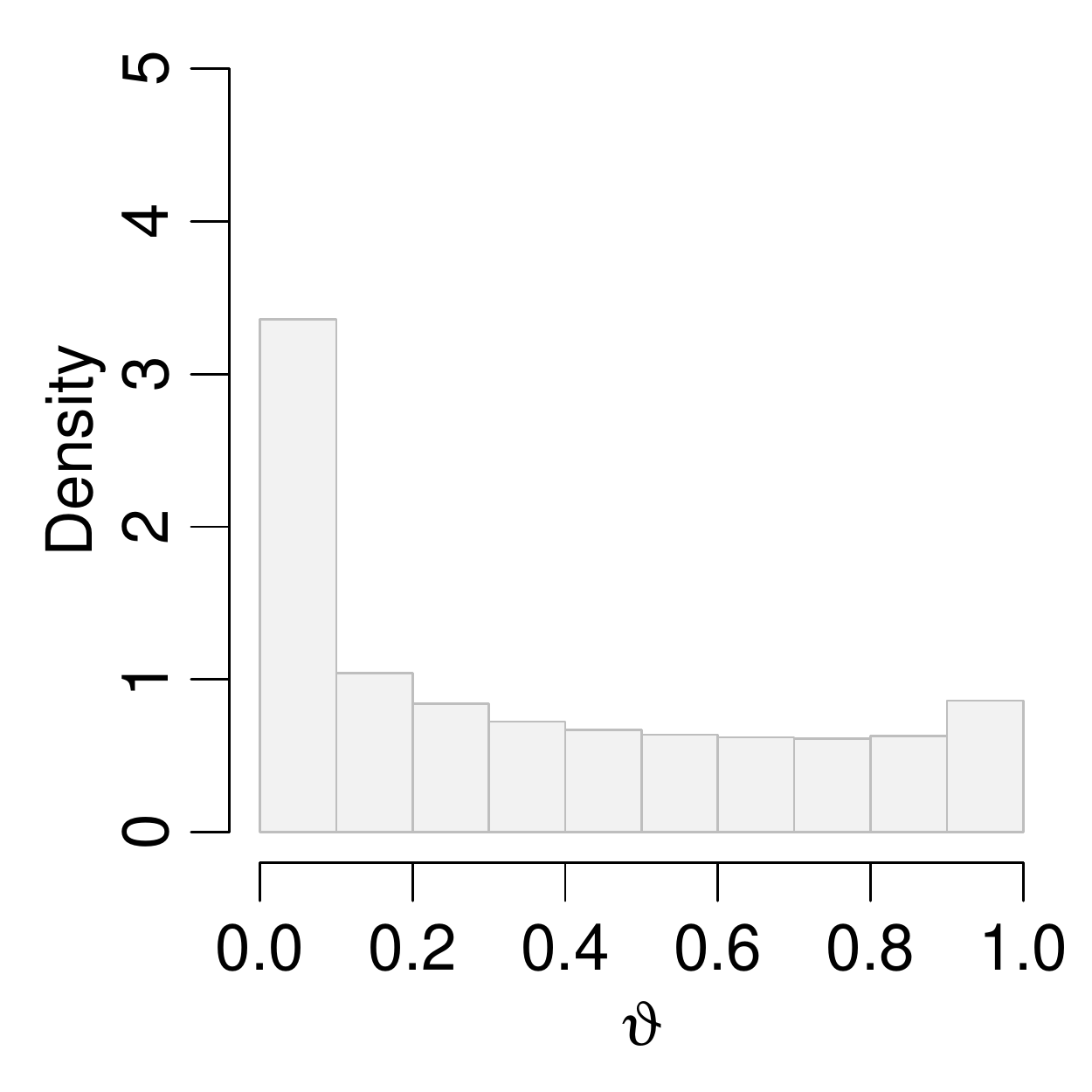}}
	\subfigure[$K=2$] {\includegraphics[scale=0.4]{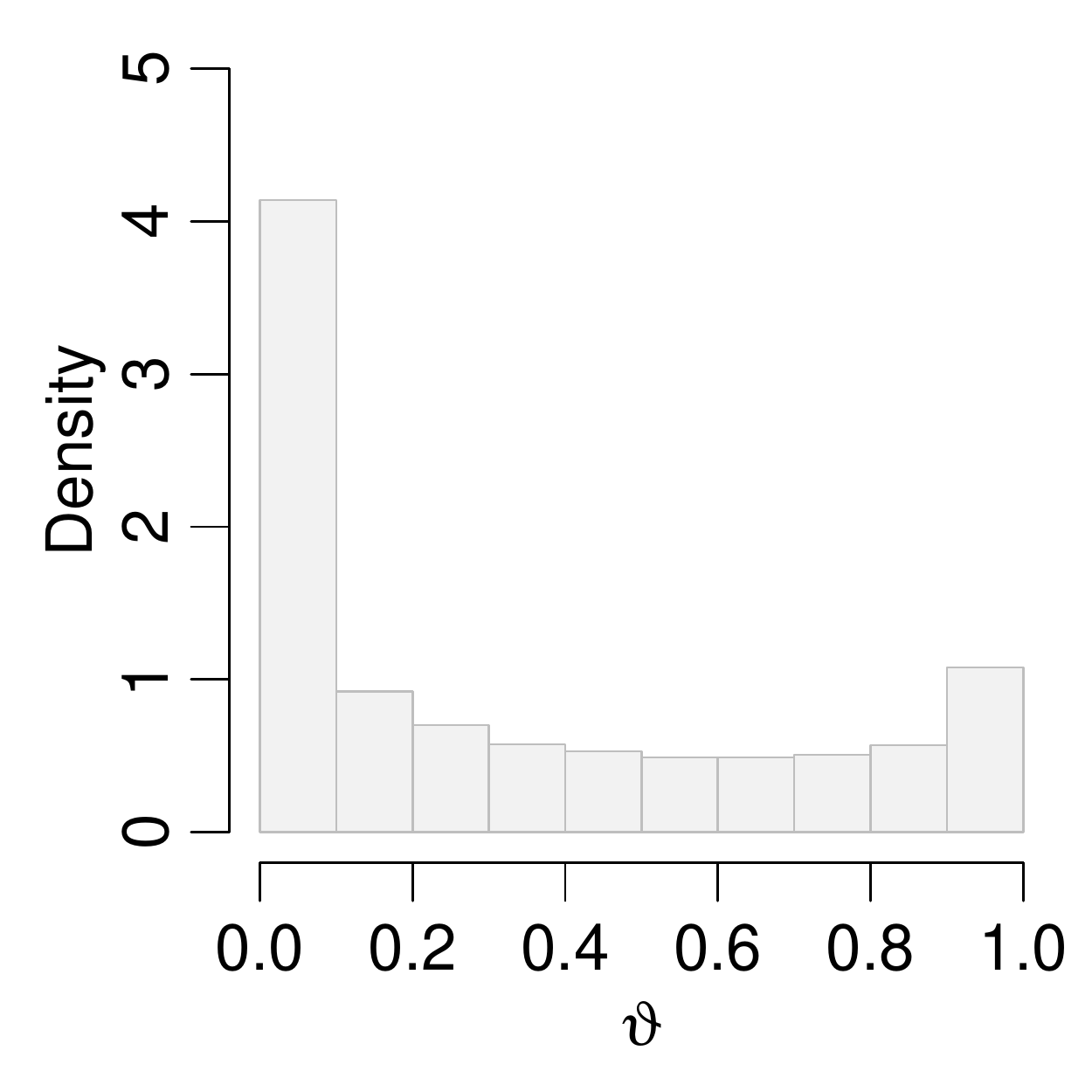}}
	\subfigure[$K=3$] {\includegraphics[scale=0.4]{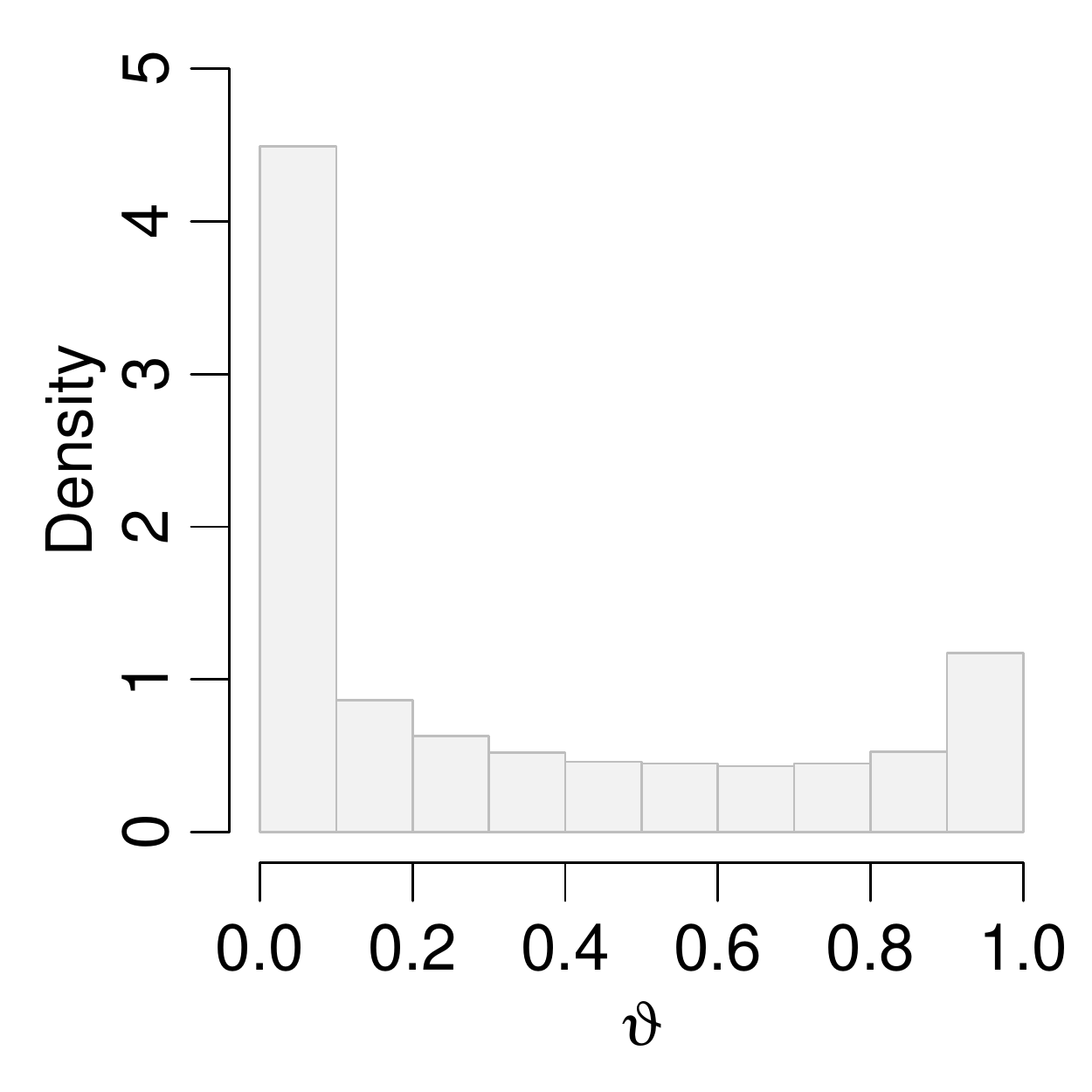}}
	\subfigure[$K=4$] {\includegraphics[scale=0.4]{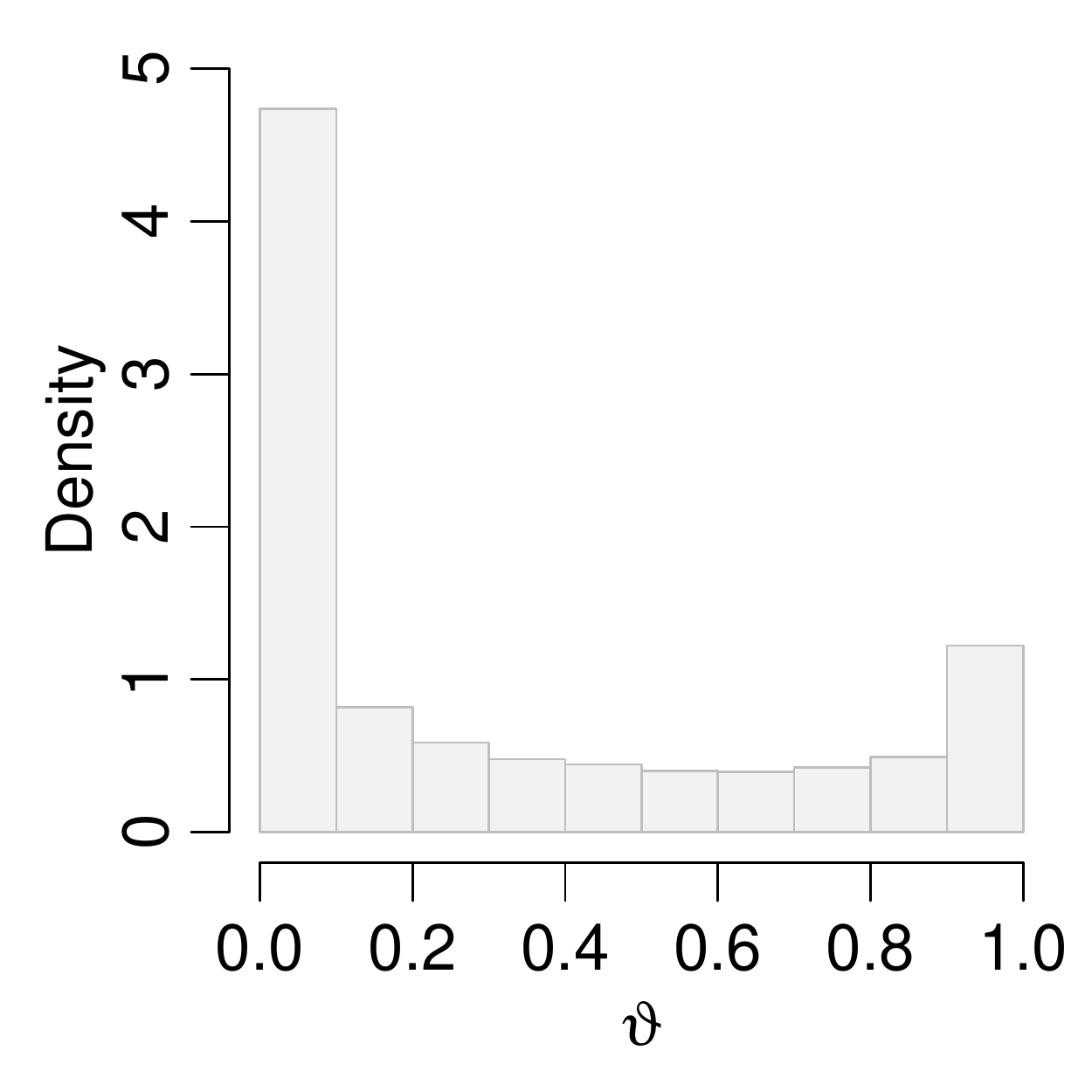}}
	\subfigure[$K=5$] {\includegraphics[scale=0.4]{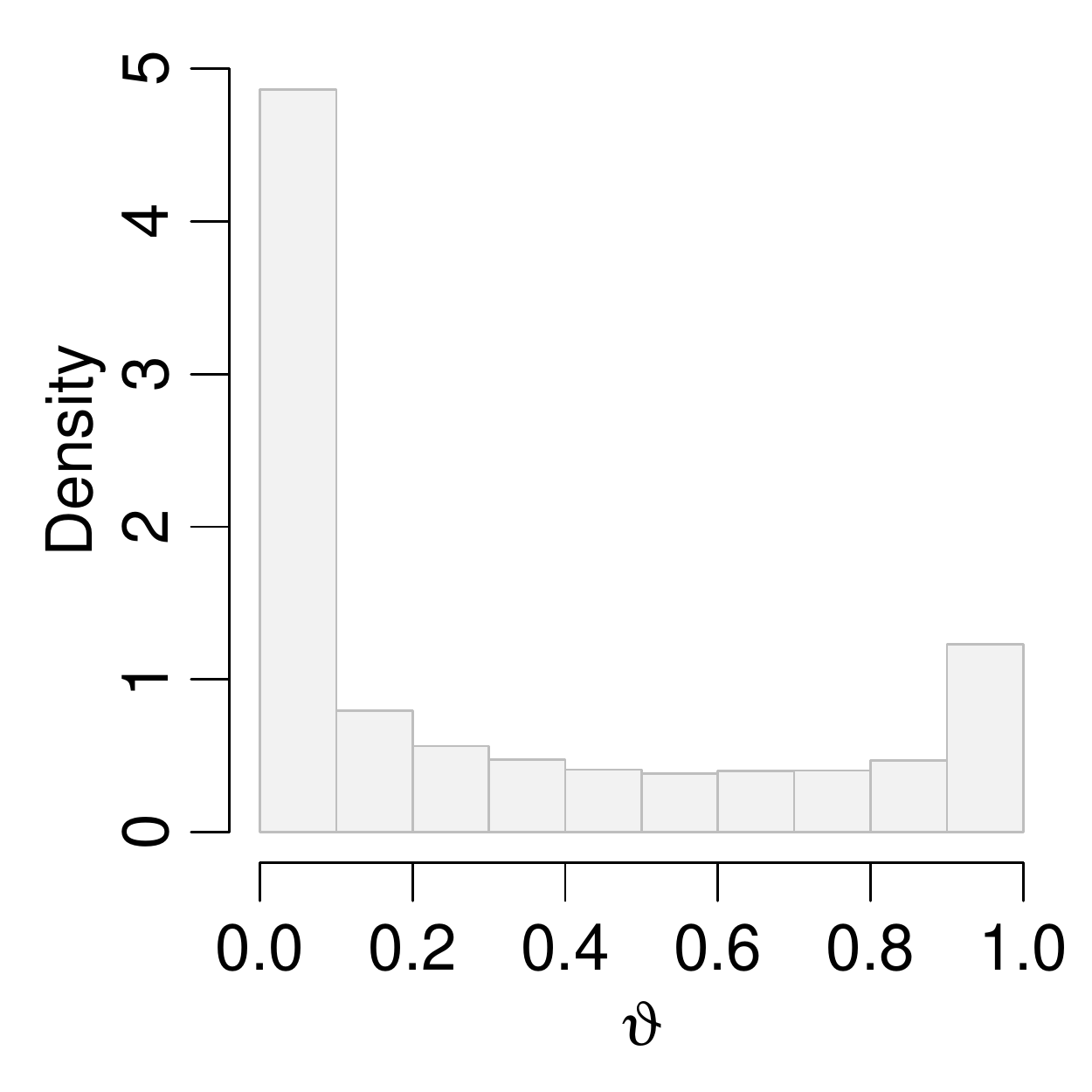}}
	\subfigure[$K=6$] {\includegraphics[scale=0.4]{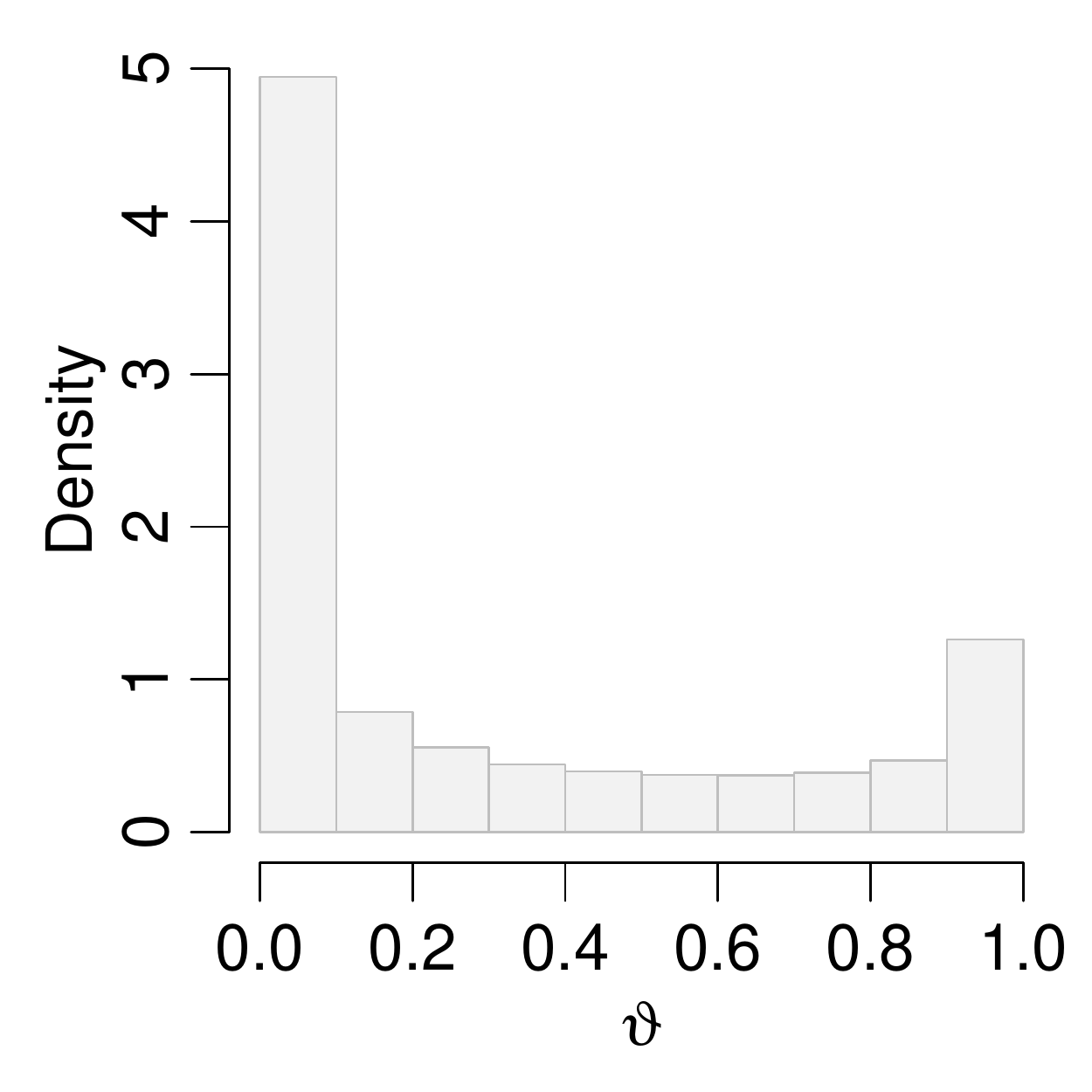}}
	\caption{Marginal prior distribution of the interaction probabilities $\vartheta_{i,i',j}$ for $K=1,\ldots,6$.}
	\label{fig_prior_simulation}
\end{figure}

Table \ref{tab_prior_especification} displays specific hyperparameter values for  $K=1, \ldots, 6$. In addition, Figure \ref{fig_prior_simulation} shows histograms of 10,000 independent realizations from the induced marginal prior distribution of the interaction probabilities, $\vartheta_{i,i',j}$, for several values of $K$. Note that these distributions are quite similar, exhibiting a mode at $\vartheta_{i,i',j} = 0.1$ (as expected) and then a somewhat uniform behavior with a slight peak towards $\vartheta_{i,i',j} = 1$.

\begin{table}[!h]
	\centering
	\begin{tabular}{cccccccccc}
		\hline
		$K$ & $\expec{d_{i,i',j}}$ & $\sd{d_{i,i',j}}$ & $b_\zeta$ & $b_\theta$ & $b_\sig$ & $b_\kap$ & $v_\zeta$ & $v_\theta$ & $v_\nu$ \\ 
		\hline
		1 & 0.377 & 0.286 & 1.508 & 2.448 & 0.074 & 0.074 & 0.868 & 1.106 & 0.192 \\ 
		2 & 0.592 & 0.308 & 2.367 & 1.546 & 0.074 & 0.074 & 1.088 & 0.879 & 0.192 \\ 
		3 & 0.752 & 0.317 & 3.009 & 1.066 & 0.074 & 0.074 & 1.227 & 0.730 & 0.192 \\ 
		4 & 0.885 & 0.321 & 3.541 & 0.740 & 0.074 & 0.074 & 1.331 & 0.608 & 0.192 \\ 
		5 & 1.003 & 0.323 & 4.010 & 0.491 & 0.074 & 0.074 & 1.416 & 0.496 & 0.192 \\ 
		6 & 1.109 & 0.327 & 4.434 & 0.290 & 0.074 & 0.074 & 1.489 & 0.381 & 0.192 \\ 
		\hline
	\end{tabular}
	\caption{Expected value and standard deviation of the prior latent distance $d_{i,i',j} = \|\,\uv_{i,j} - \uv_{i',j}\|\,$, rate hyperparameters, and variance components for the prior distributions on the variance parameters for $K=1,\ldots,6$.} 
	\label{tab_prior_especification} 
\end{table}

\section{Computation} \label{sec_computation}

For a given latent dimension $K$, the posterior distribution $p(\UPS\mid\Y)$ is explored using Markov chain Monte Carlo methods (MCMC; e.g., \citealp{gamerman2006markov}). The computational algorithm entails a combination of Gibbs sampling and Metropolis-Hastings steps (e.g., \citealp{haario2001adaptive}). Details about the MCMC algorithm can be found in the Appendix \ref{app_mcmc}. In the following sections we discuss the issues of identifiability and model selection which are common in the latent space modeling framework.



\subsection{Identifiability}\label{sec_identifiability}

Our proposed MNLPM inherits the property of invariance to rotations and reflections of the social space from the simple latent space model of \cite{hoff-2002}. Indeed, for any $K\times K$ orthogonal matrix $\Q$, the likelihood associated with the reparameterization $\tilde{\uv}_{i,j} = \Q \uv_{i,j}$ is independent of $\Q$ since $\|\, \uv_{i,j} - \uv_{i',j} \,\|$ = $\|\, \tilde\uv_{i,j} - \tilde\uv_{i',j} \,\|$.  This lack of identifiability does not affect our ability to make inferences on the interaction probabilities $\vartheta_{i,i',j}$\,s (which are identifiable). However, it does hinder our ability to provide posterior estimates of latent-position based measures (e.g., network correlation), including the latent positions themselves.

We address this invariance issue using a common parameter expansion used by \cite{hoff-2002} and many others. In particular, the problem of identifiability is addressed through a post-processing step in which the $B$ posterior samples are rotated/reflected to a shared coordinate system.  For each sample $\UPS^{(b)}$, for $b=1, \ldots, B$, an orthogonal transformation matrix $\Q^{(b)}$ is obtained by minimizing the Procrustes distance,
\begin{align}\label{eq:proc}
\tilde{\Q}^{(b)} = \argmin_{\Q \in \mathcal{S}^{K}} \tr\left\{ \left( \E^{(1)}-\E^{(b)}\Q \right)^\trans \left( \E^{(1)}-\E^{(b)}\Q \right) \right\}\,,
\end{align}
where $\mathcal{S}^{K}$ denotes the set of $K \times K$ orthogonal matrices and $\E^{(b)}$ is the $I \times K$ matrix whose $I$ rows correspond to the transpose of $\etav_1^{(b)}, \ldots, \etav_I^{(b)}$. The minimization problem in \eqref{eq:proc} can be easily solved using singular value decompositions (e.g., see \citealp{borg-2005}). Indeed, $\tilde{\Q}^{(b)} = \R^{(b)}\L^{(b)\,\trans}$, where $\L^{(b)}\D^{(b)}\R^{(b)\,\trans}$ is the singular value decomposition of $\E^{(1)\,\trans}\E^{(b)}$. Once the matrices $\tilde{\Q}^{(1)}, \ldots, \tilde{\Q}^{(B)}$ have been obtained, posterior inference for the latent positions of our model is based on the transformed coordinates $\tilde{\uv}_{i,j}^{(b)} = \tilde{\Q}^{(b)} \uv_{i,j}^{(b)}$ and $\tilde{\etav}^{(b)}_i = \tilde{\Q}^{(b)} \etav^{(b)}_i$.

\subsection{Model selection}

In general, setting $K = 2$ for the dimension of the latent space is a sensible choice, as it simplifies the visualization and description of social relationships. However, our objective goes beyond a mere description of multilayer network data, an in consequence the value of $K$ plays a critical role in the results.

The network literature has largely focused on the Bayesian Information Criterion (BIC; e.g., \citealp{hoff-2005}, \citealp{handcock-2007}, \citealp{airoldi-2009}). However, the BIC is often inappropriate for hierarchical models since the hierarchical structure implies that the effective number of parameters will be typically less than the actual number of parameters in the likelihood. An alternative to the BIC is the Watanabe-Akaike Information Criterion (\textsf{WAIC}; \citealp{watanabe2010asymptotic}, \citealp{watanabe2013widely}, \citealp{gelman2014understanding}),
$$
\textsf{WAIC} \left(K\right) = - 2\sum_j\sum_{i,i':i < i'} \log\textsf{E}\left[ p\left( y_{i, i', j} \mid  \mathbf{\Upsilon} \right)\right] + 2p_{\text{WAIC}}\,,
$$
where 
$$
p_{\text{WAIC}} = 2 \sum_j\sum_{i,i':i<i'} \left( \log \textsf{E} \left[ p \left( y_{i, i', j} \mid  \mathbf\Upsilon \right) \right]   - \textsf{E}\left[\log p \left(  y_{i, i', j} \mid  \mathbf\Upsilon  \right) \right] \right)
$$
is the model complexity (effective number of parameters), and $\UPS$ is the set of model parameters assuming that the dimension of the social space is $K$. The expected values in these expressions are calculated with respect to the posterior distribution $p(\UPS\mid\Y)$, and can be approximated by averaging over the MCMC samples $\UPS^{(1)},\ldots,\UPS^{(B)}$.

\section{Illustrations}\label{sec_ilustrations}

In this section, we illustrate and evaluate the performance of our MNLPM using two benchmark data sets: The bank wiring room data of \cite{roethlisberger2003management} and the friendship cognitive social structure of \citet{krackhardt-1987}. The main contributions involving the characterization of the social dynamics with our MNLPM are formal approaches to measure network correlation and perceptual agreement, which are respectively illustrated  with each data set. 

\subsection{Bank wiring room data}

\begin{figure}[!h]
	\centering
	\subfigure[Horseplay] {\includegraphics[scale=0.15]{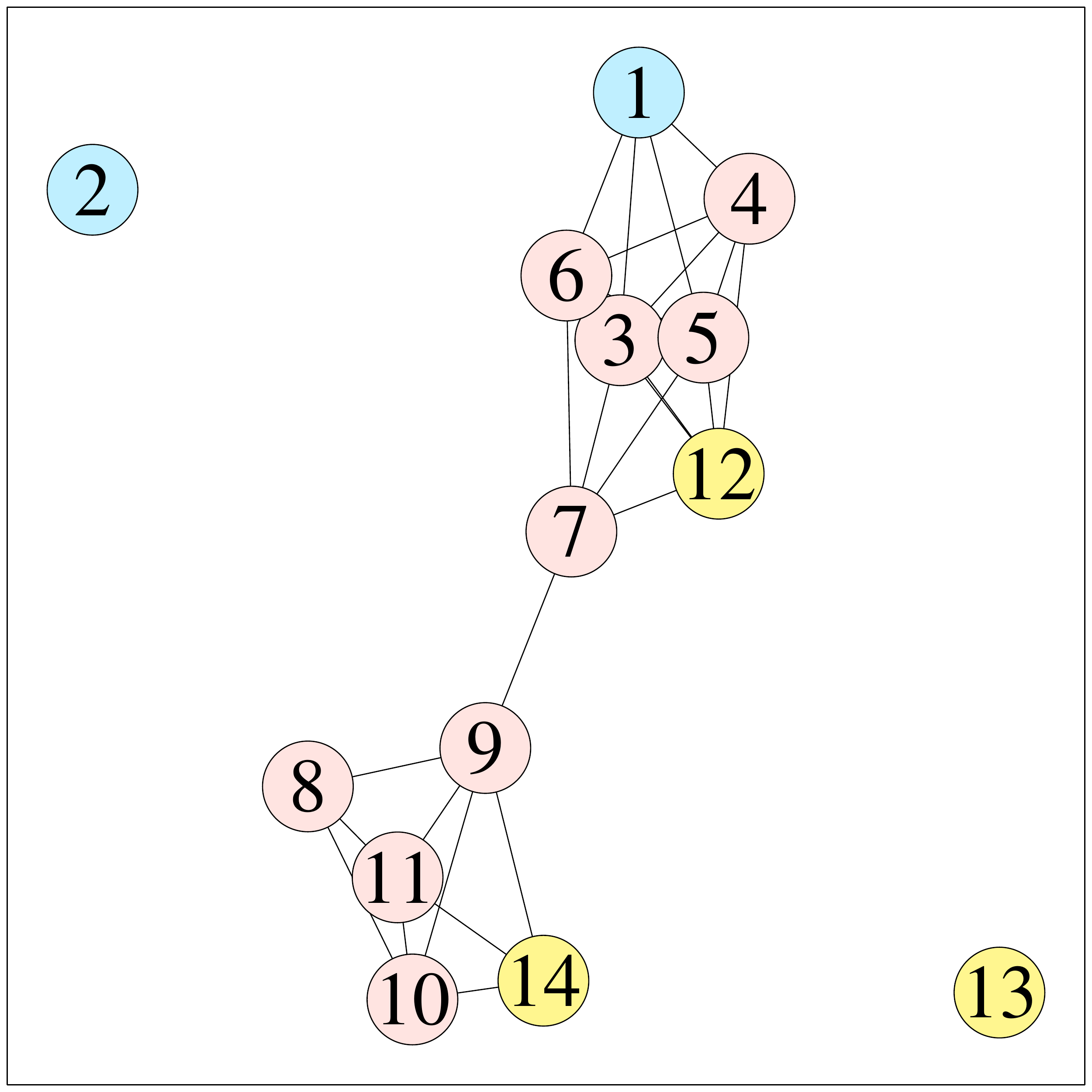}}
	\subfigure[Arguments] {\includegraphics[scale=0.15]{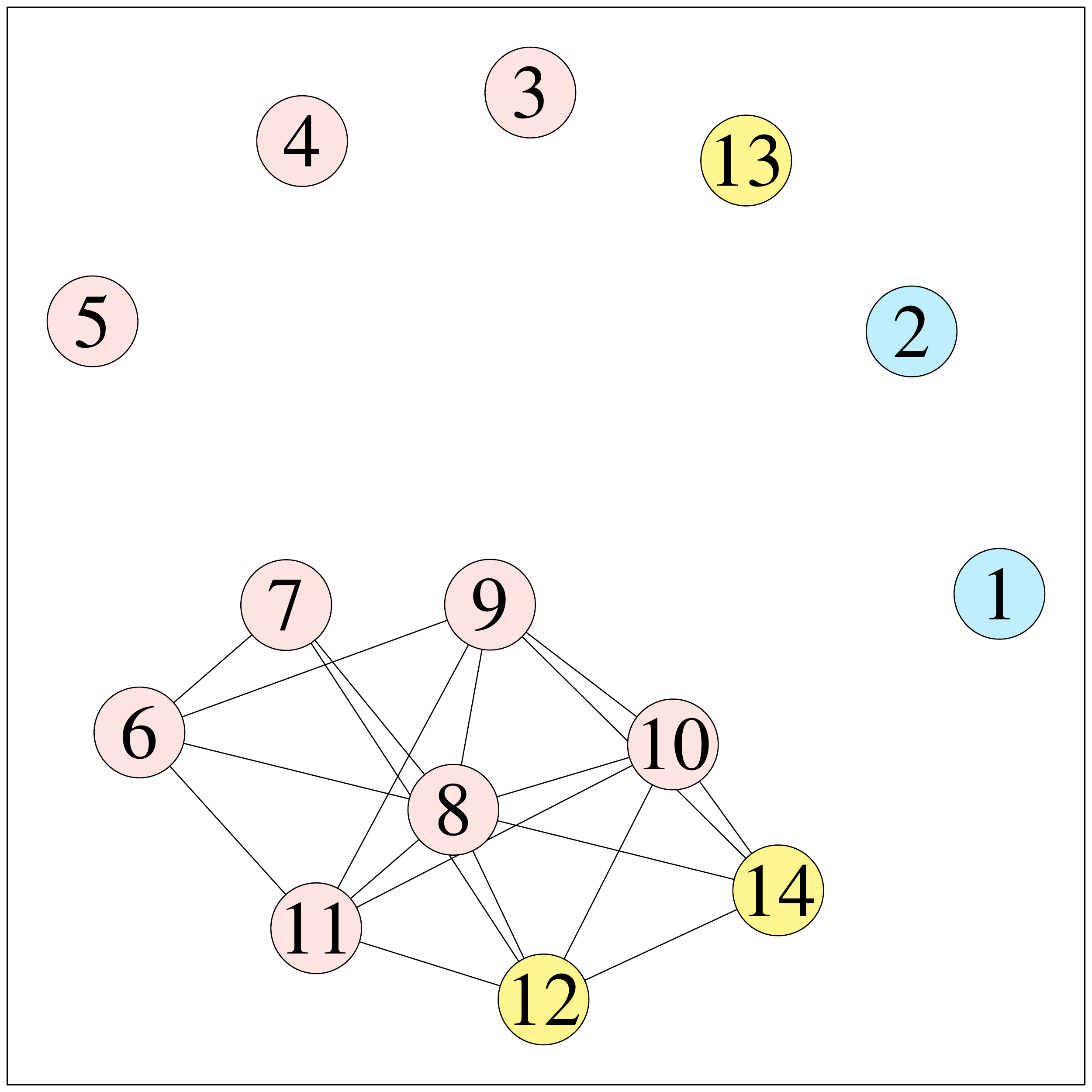}}
	\subfigure[Frienship] {\includegraphics[scale=0.15]{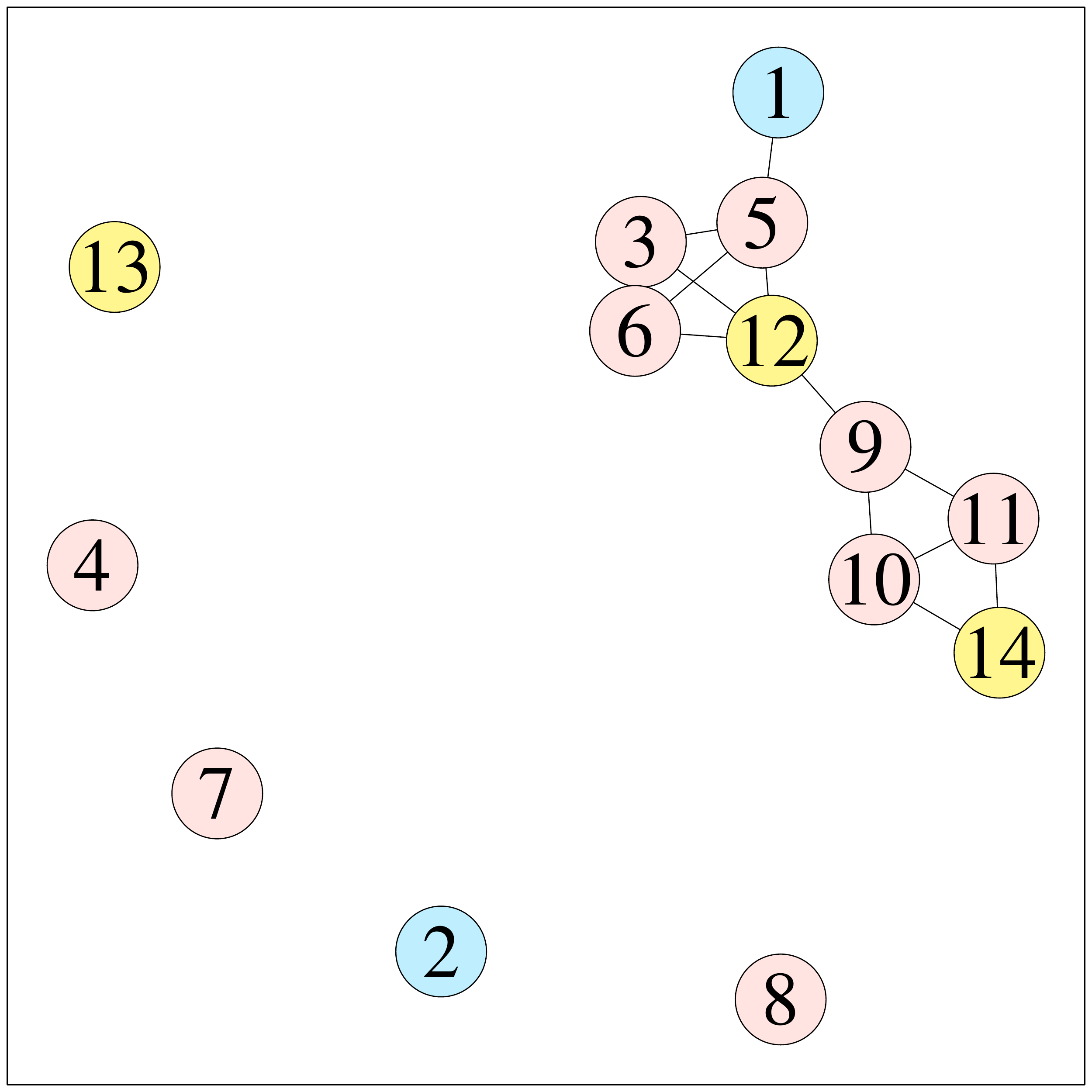}}
	\subfigure[Antagonism]{\includegraphics[scale=0.15]{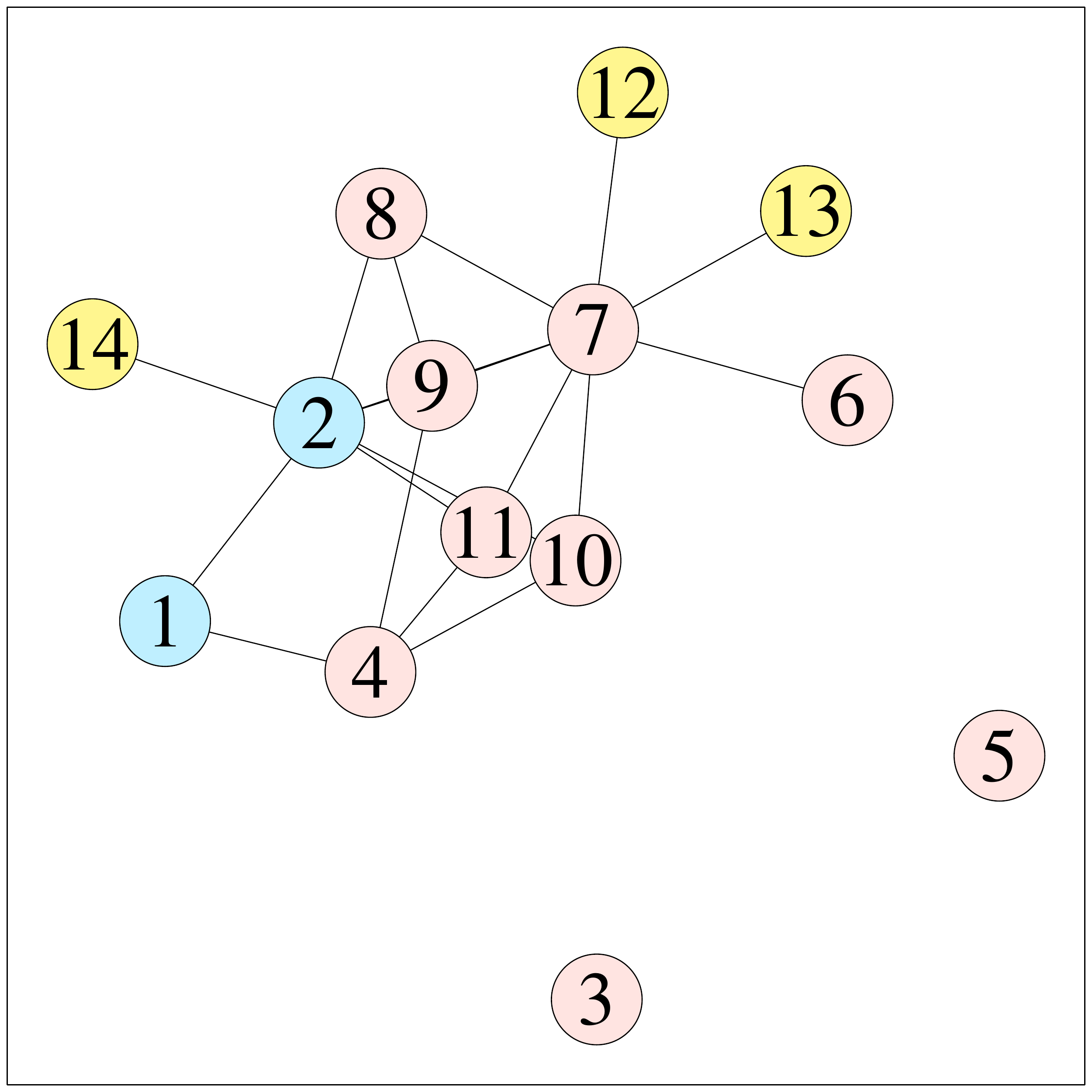}}
	\caption{Visualization of the bank wiring room data. Inspectors are shown in blue, solderers in yellow, and wiremen or assemblers in red.}
	\label{fig_graphs_wiring}
\end{figure}

These are the observational data on $I=14$ Western Electric (Hawthorne Plant) employees from the bank wiring room presented in \cite{roethlisberger2003management}. The employees worked in a single room and include two inspectors (actors 1 and 2), three solderers (actors 12, 13, and 14), and nine wiremen or assemblers (actors 3 to 11). The authors gathered data about $J = 4$ symmetric interaction categories including: participation in horseplay (Horseplay, network 1), participation in arguments about open windows (Arguments, network 2), friendship (Friendship, network 3), and antagonistic behaviour (Antagonism, network 4). This dataset is considered nowadays as a standard referent to test models for multilayer network data (e.g., \citealp{bartz2014parallel}, \citealp{liu202010th}, and \citealp{abdollahpouri2020multi}). Figure \ref{fig_graphs_wiring} displays a visualization of all the relational layers.

\begin{figure}[!h]
	\centering
	\subfigure[Information criterion] {\includegraphics[scale=0.76]{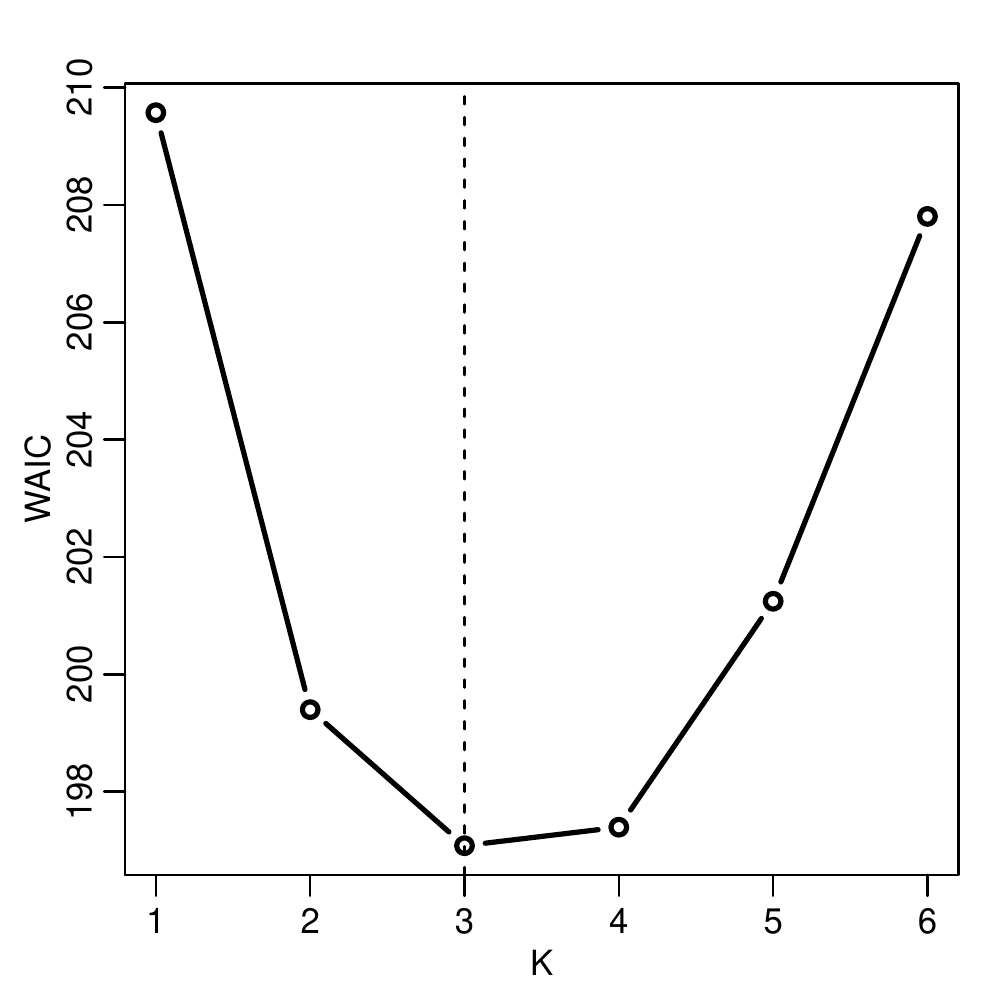}}
	\subfigure[Log-likelihood, $K=3$] {\includegraphics[scale=0.76]{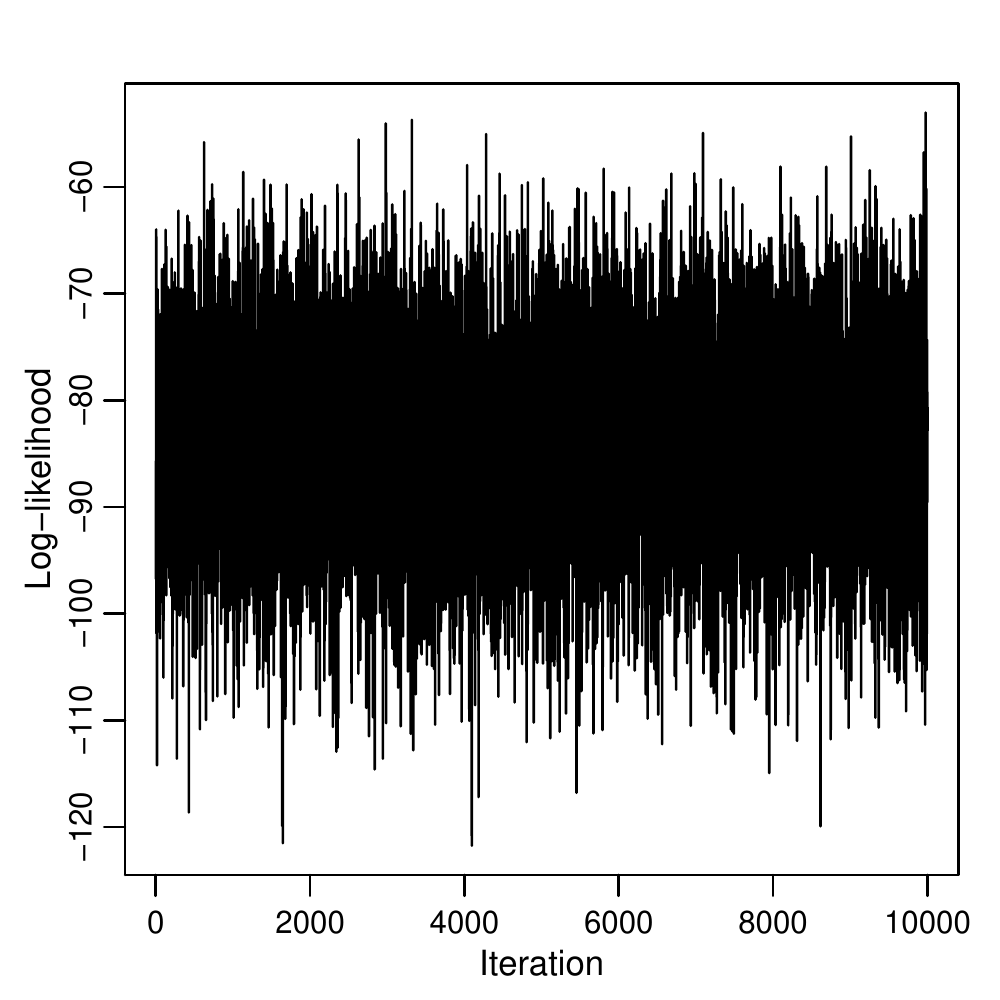}}
	\caption{\textsf{WAIC} values to select the latent dimension $K$ of the social space for the Bayesian analysis of the bank wiring room data using our MNLPM, and log-likelihood chain associated with the value of $K$ optimizes the \textsf{WAIC} ($K=3$).} 
	\label{fig_WAIC_ll_wiring}
\end{figure}

We implement our MNLPM for the bank wiring room data following the computational approach described in Section \ref{sec_computation}. The results presented below are based on $B = 10,000$ samples of the posterior distribution obtained after thinning the original Markov chains every 10 observations and a burn-in period of 100,000 iterations. All chains mix reasonably well. Left panel of Figure \ref{fig_WAIC_ll_wiring} shows the \textsf{WAIC} computed for several MNLPMs fitted using a range of latent dimensions of the social space. This criterion clearly supports $K = 3$ (\textsf{WAIC} = 197.1) as an optimal choice, which is the latent dimension we use in all our analyses henceforth. The effective sample sizes of the model parameters following the MCMC algorithm discussed above range from 4,081 to 10,000. Right panel of Figure \ref{fig_WAIC_ll_wiring} displays the log-likelihood chain associated with the latent dimension that optimizes the \textsf{WAIC}, which shows no signs of lack of convergence.

\subsubsection{Consensus network}

\begin{figure}[!h]
	\centering
	\includegraphics[scale=0.52]{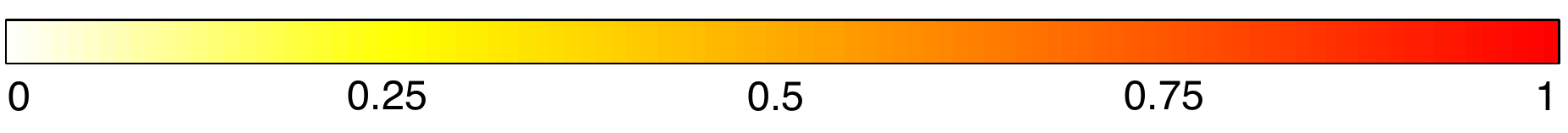}
	\subfigure[$\expec{\upsilon_{i,i'} \mid \Y}$] {\includegraphics[scale=0.76]{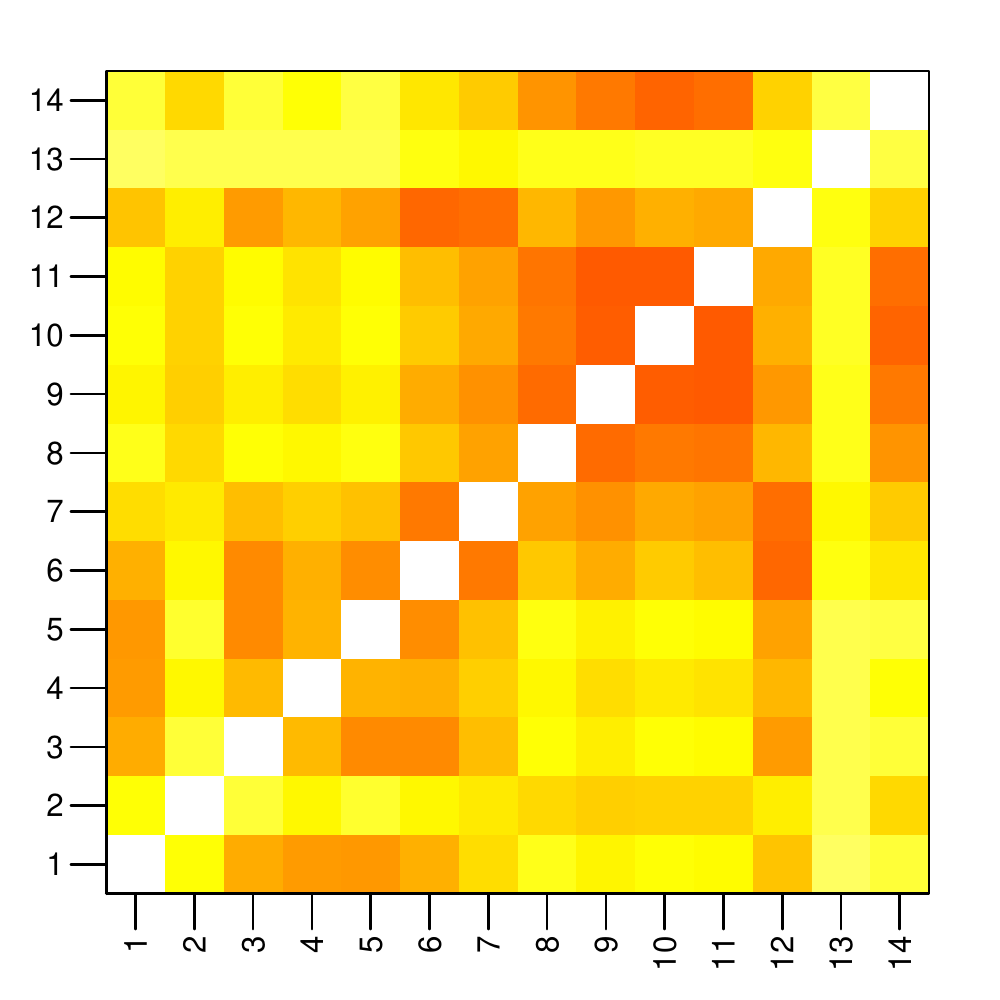}}
	\subfigure[$\frac{1}{J} \sum_{j=1}^{I} y_{i,i',j}$] {\includegraphics[scale=0.76]{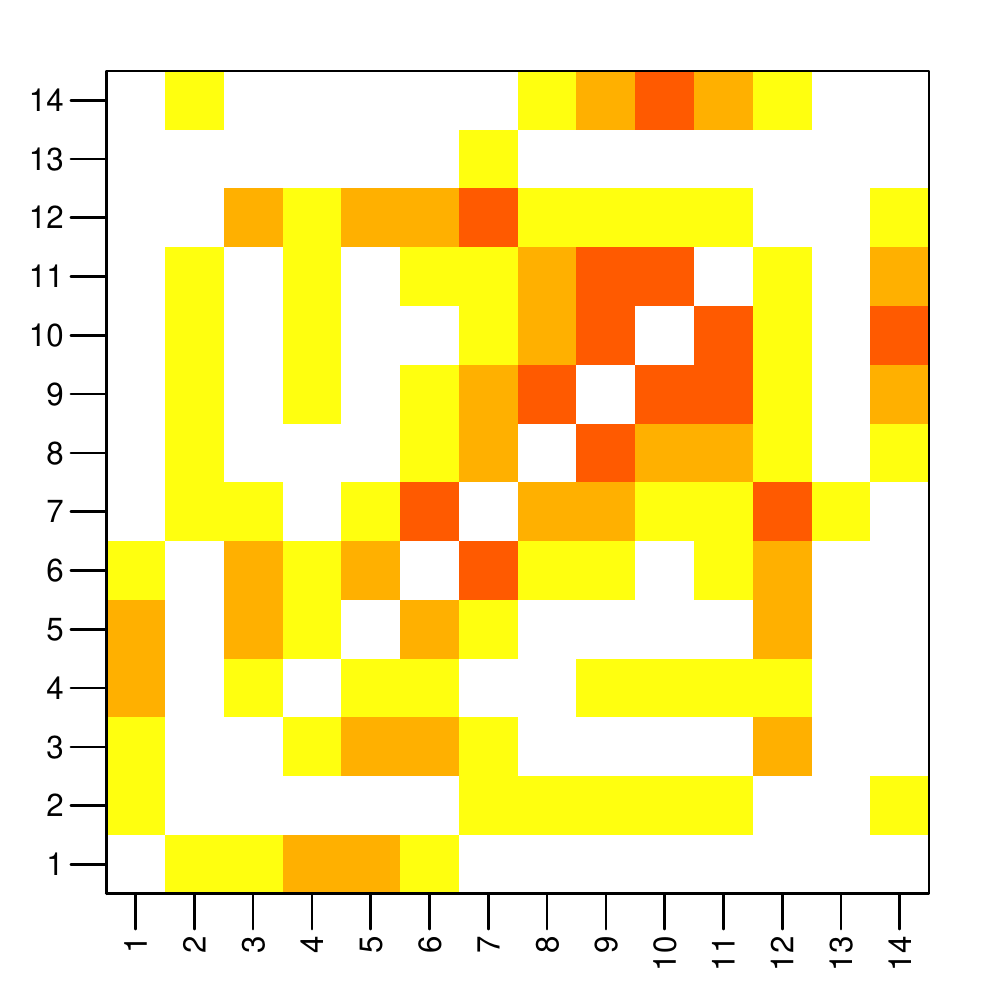}}
	\caption{Consensus network estimates for the bank wiring room data. The left panel provides the posterior mean under our MNLPM, and the right panel shows the proportion of observed links across networks.}\label{fig_post_probs_concensus_wiring}
\end{figure}

Unlike other models for multilayer network data (e.g., \citealp{salter-2017}), our approach provides a straightforward mechanism to construct a ``consensus'' network. Indeed, the average positions $\etav_1,\ldots,\etav_I$ can be used to generate a weighted network, $\upsilon_{i,i'} = \Phi\left(\mu_\zeta - e^{\mu_\theta}\|\, \etav_i - \etav_{i'} \|\, \right)$, that ``collapses'' all the relational layers in a single network by weighting them according to the mean parameters of the hierarchical prior distribution. The consensus network can be very useful when an overall summary of the social dynamics is required. Figure \ref{fig_post_probs_concensus_wiring} displays heat-maps for the matrix of posterior means $\expec{\upsilon_{i,i'} \mid \Y}$
and the proportion of observed links across networks, $\frac{1}{J} \sum_{j=1}^{I} y_{i,i',j}$. Although the estimate provided by our model is ``denser'' than the empirical proportion, note that the estimates are  similar. This also suggests that the model correctly characterizes the data generating process.

\subsubsection{Projections in social space}

\begin{figure}[!h]
	\centering
	\subfigure[Horseplay] {\includegraphics[scale=0.15]{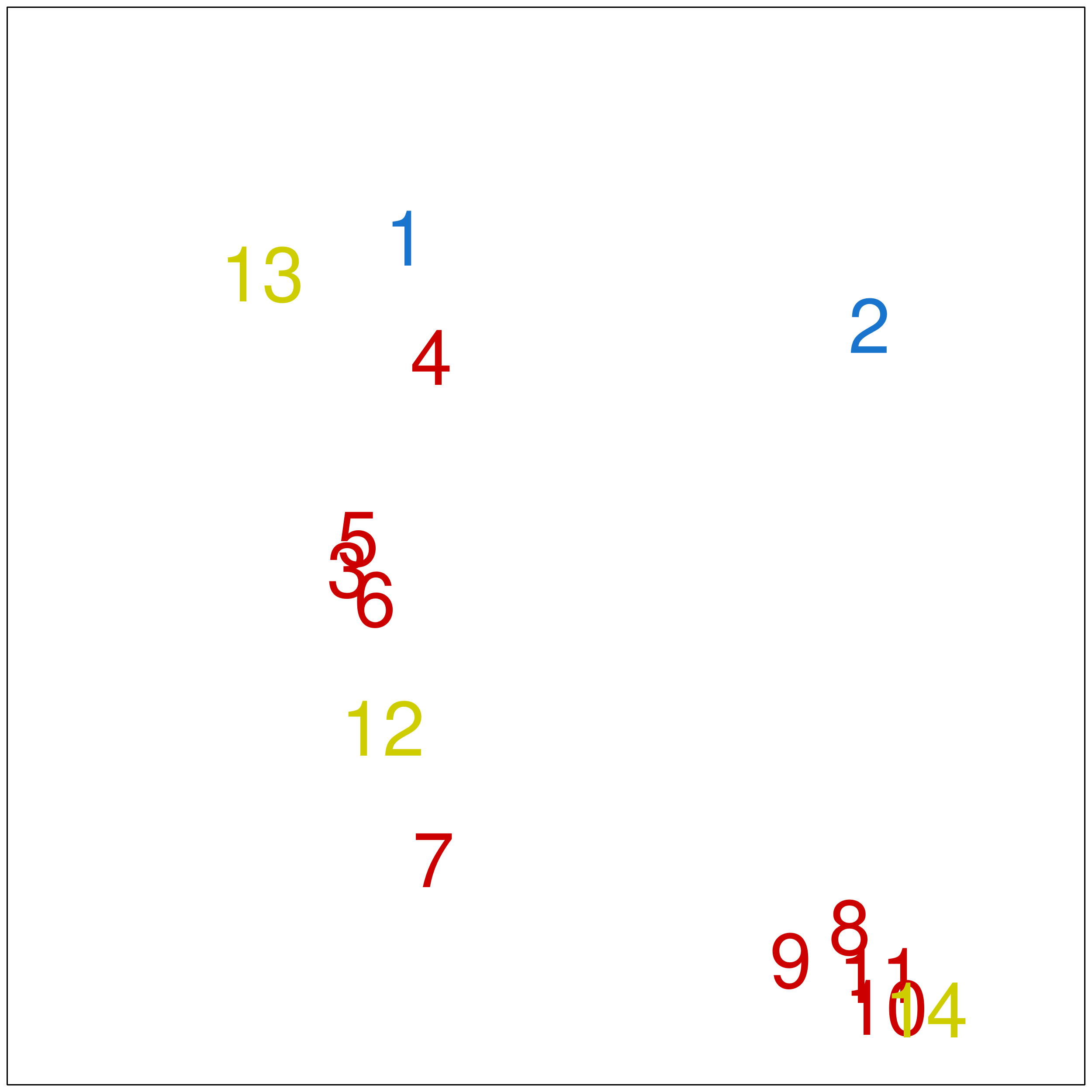}}
	\subfigure[Arguments] {\includegraphics[scale=0.15]{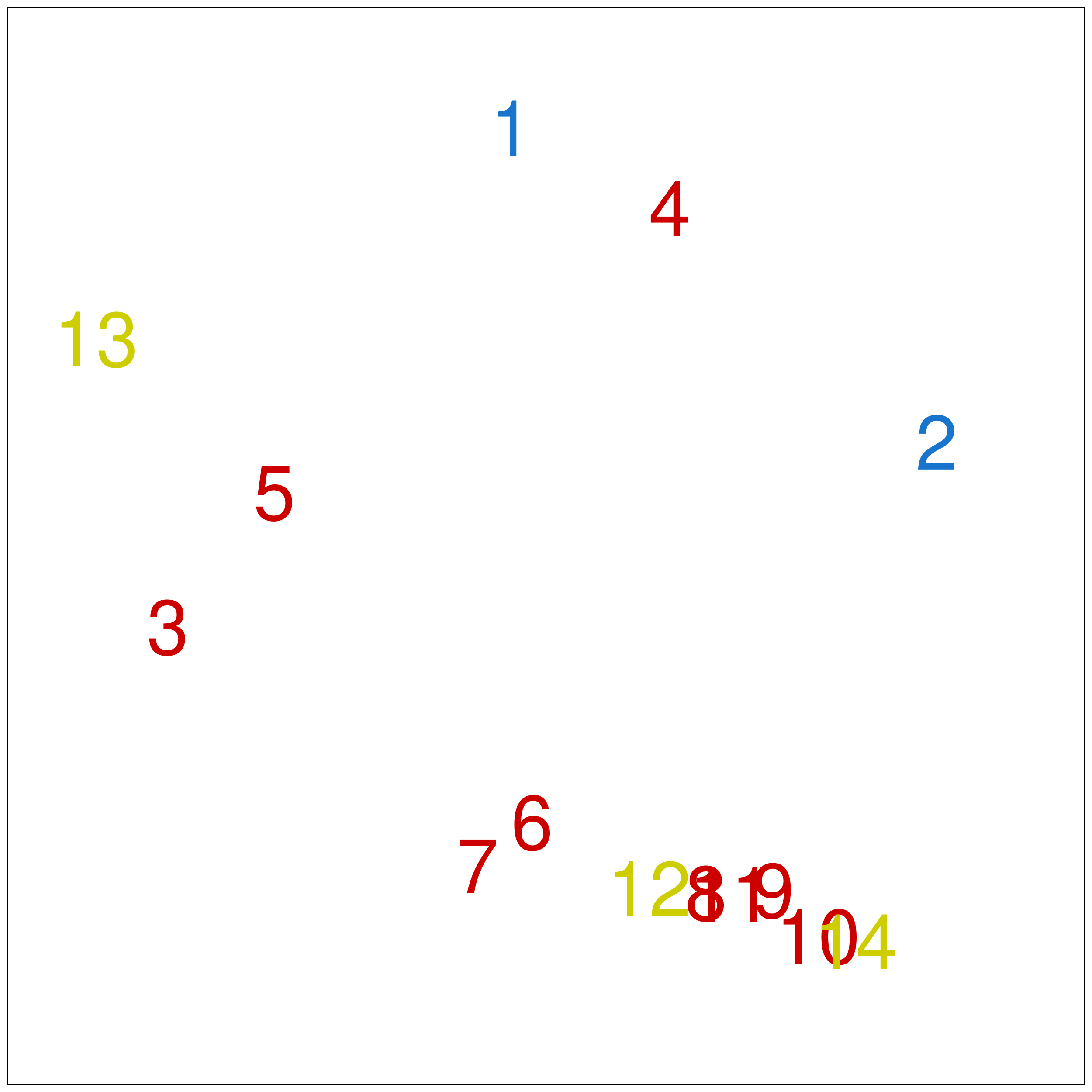}}
	\subfigure[Frienship] {\includegraphics[scale=0.15]{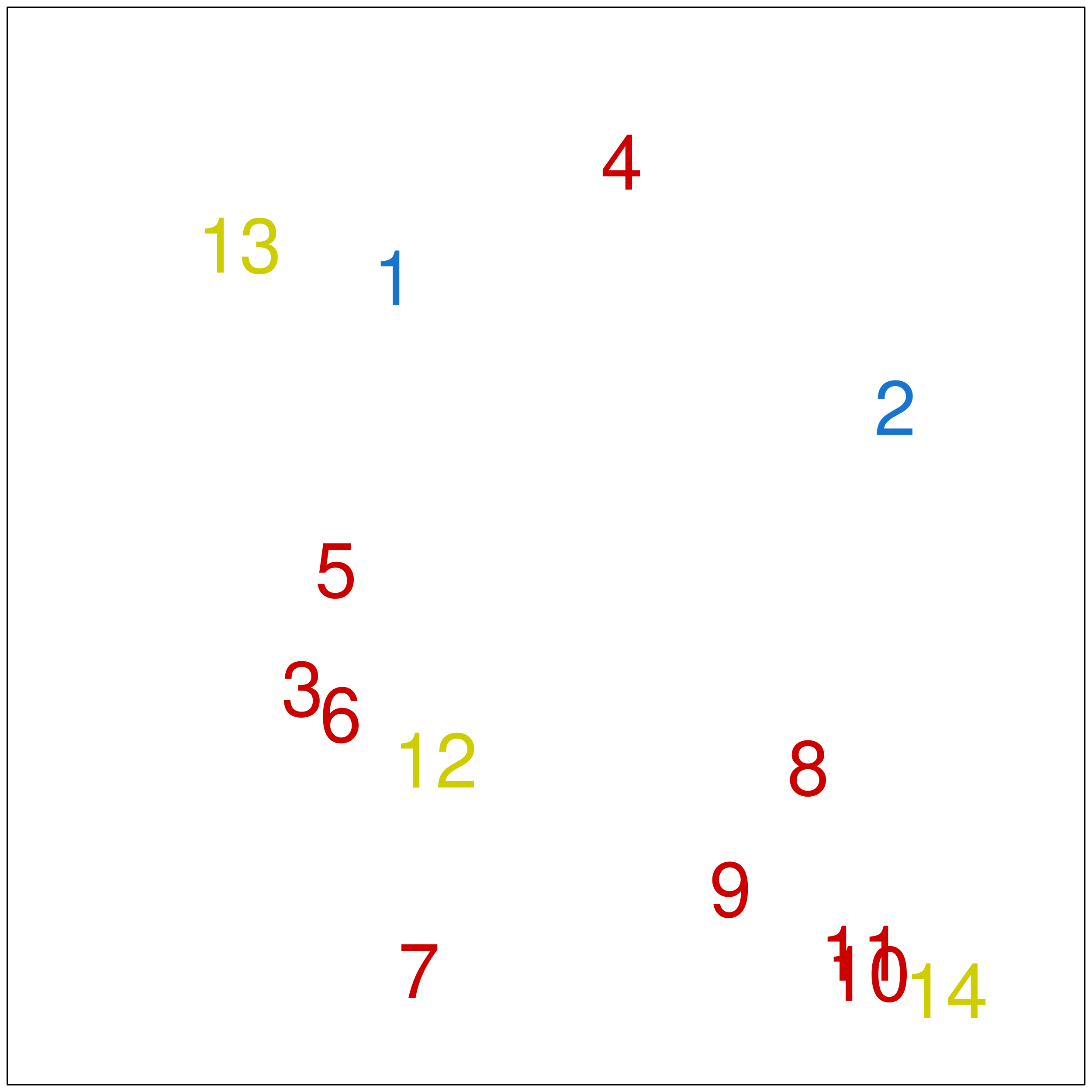}}
	\subfigure[Antagonism]{\includegraphics[scale=0.15]{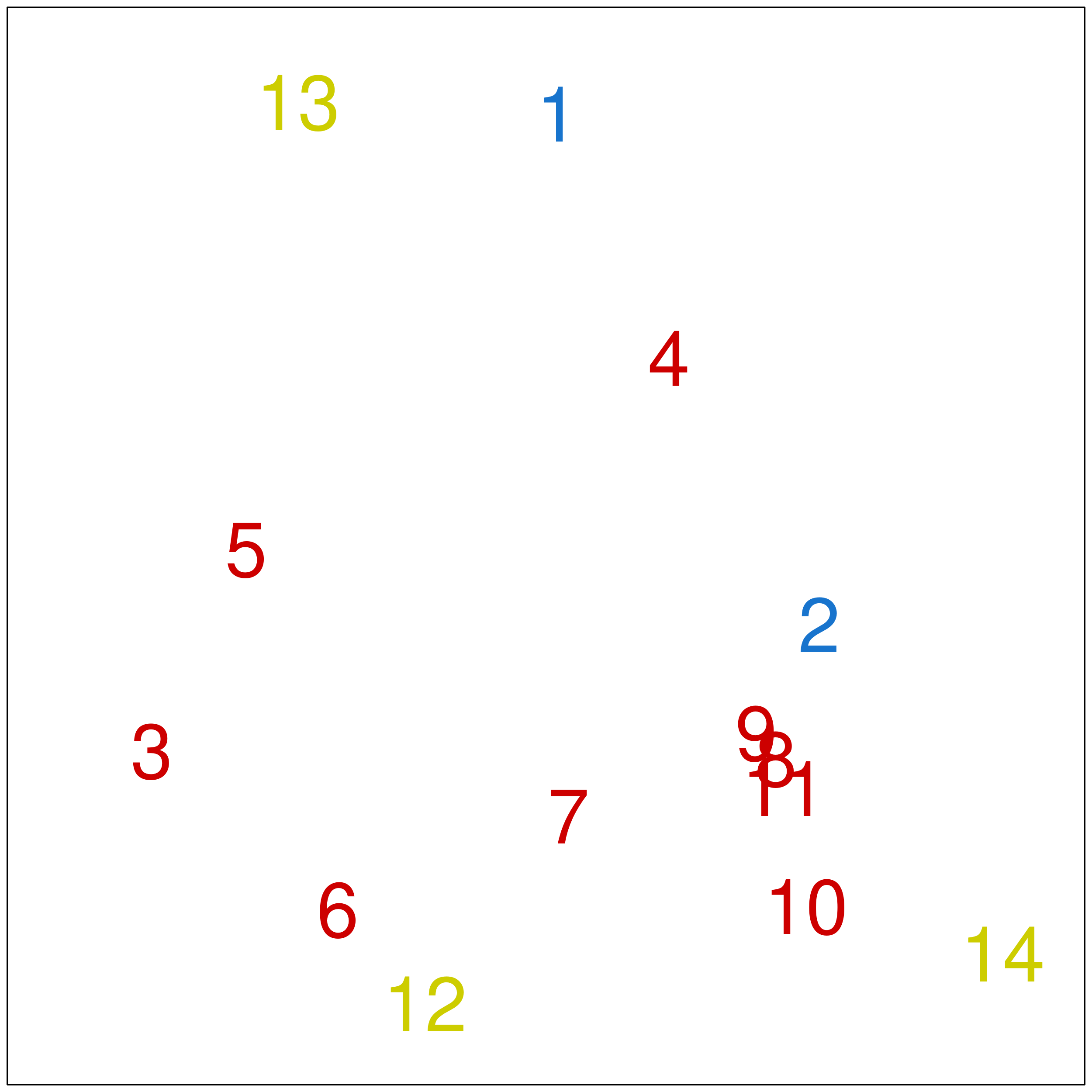}}
	\caption{Posterior means of Procrustes-transformed latent positions $\expec{\tilde\uv_{i,j}\mid\Y}$ along the two dimensions with the highest variability, for the bank wiring room data. Inspectors are shown in blue, solderers in yellow, and wiremen or assemblers in red.} \label{fig_latent_positions_u_wiring}
\end{figure}

Actor-specific latent positions $\uv_{1,1},\ldots,\uv_{I,J}$ provide a powerful tool for describing social interactions. Figure \ref{fig_latent_positions_u_wiring} shows Procrustes-transformed latent position estimates $\expec{\tilde\uv_{i,j}\mid\Y}$ along the two dimensions with the highest variability for each layer of the system. Even though the social behaviour is similar across layers, there are important particularities. First, note that the social dynamics of Horseplay and Friendship are quite similar, except that in Horseplay latent positions seem to be more clustered together. Furthermore, many social patterns are evident. For instance, actors who have close positions in Horseplay and Friendship, typically have distant positions in Antagonism; such an effect is particularly clear among actors 3 and 6, and actors 10 and 14. Lastly, complementing the insights provided by the consensus network, average positions $\etav_1,\ldots,\etav_I$ can be also very useful to perform an ``global'' visualization of the social dynamics.

As pointed out above, latent positions allow us to distinguish groups of actors that fulfil similar social roles. In order to identify such clusters, we can either apply some unsupervised clustering technique (e.g., hierarchical clustering, $k$-means clustering) or include directly into the model a set of parameters that assign actors to groups. The latter is quite preferable since the uncertainty related to the clustering task can be directly quantified along with its relationship to other model parameters (see Section \ref{sec_discussion} for more details).

\subsubsection{Network correlation}

\begin{figure}[!h]
	\centering
	\subfigure[Bank wiring room data] {\includegraphics[scale=0.76]{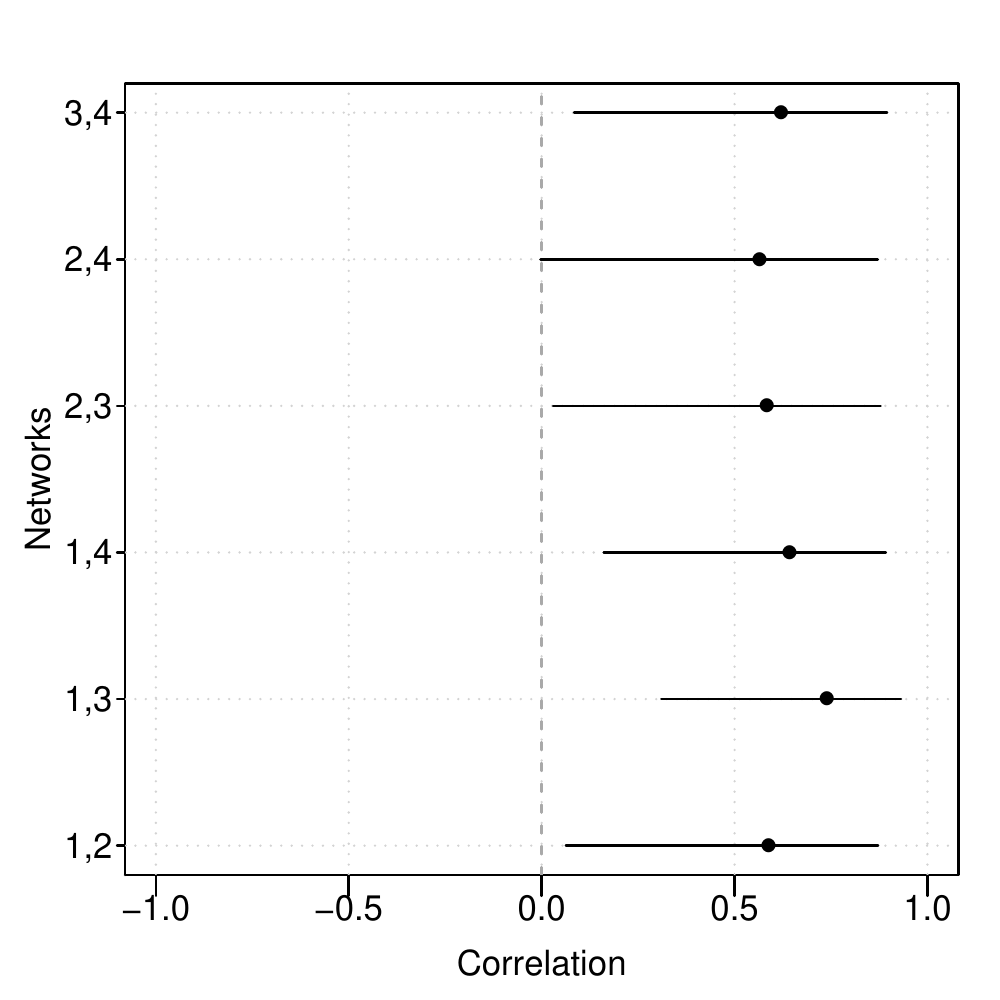}}
	\subfigure[Synthetic data]        {\includegraphics[scale=0.76]{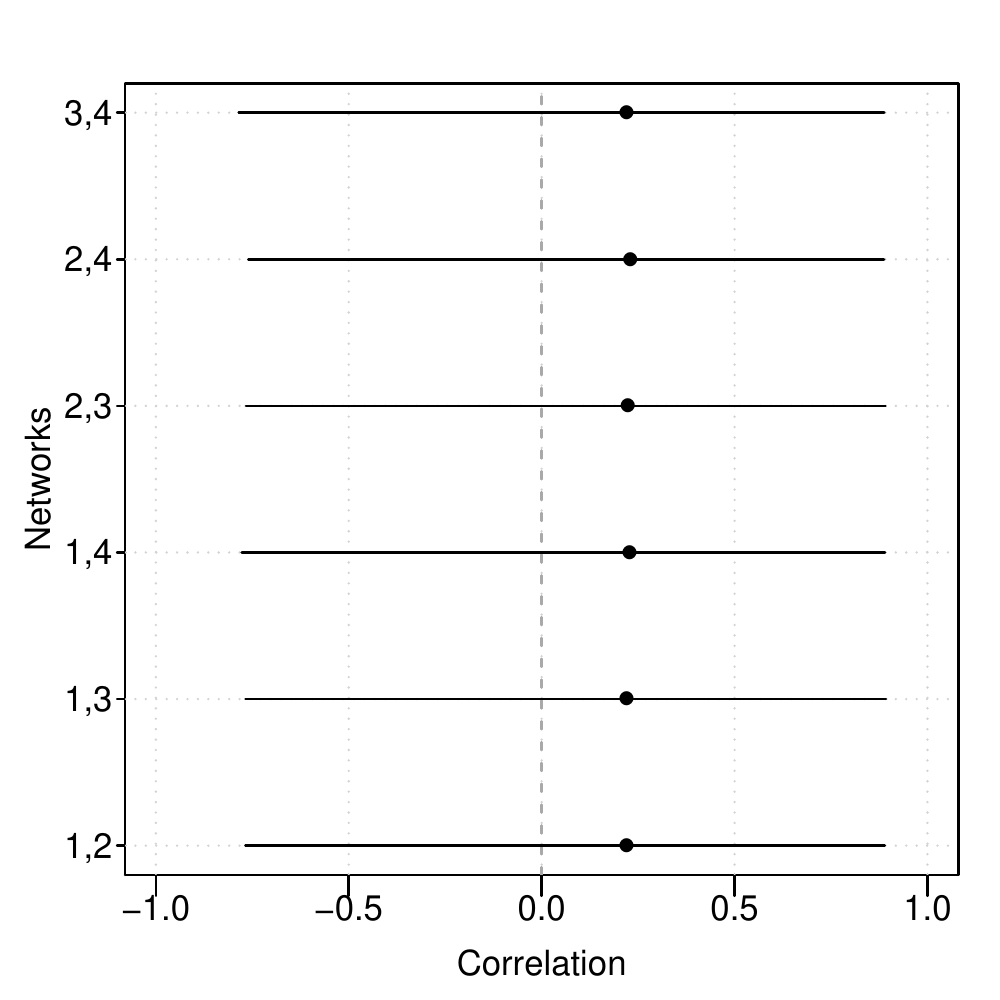}}
	\caption{95\% credible intervals and posterior means for all pairwise network correlations $\rho_{j,j'}$. Left panel: Bank wiring room data (participation in horseplay, network 1; participation in arguments about open windows, network 2; friendship, network 3; and antagonistic behaviour, network 4). Right panel: Synthetic data.}
	\label{fig_correlation_ic_wiring}
\end{figure}

Another key feature of the MNLPM is that it implicitly allows us to obtain correlation measures between layers as a direct by-product of the model parameterization. Indeed, we define the correlation between layers $j$ and $j'$, for $j,j'=1,\ldots,J$, as 
$$\rho_{j,j'} = \textsf{cor}(u^*_{1,j},\ldots, u^*_{I,j},u^*_{1,j'},\ldots, u^*_{I,j'})\,,$$
where $u^*_{i,j}$ is the maximum Procrustes-transformed latent characteristic across latent dimensions of actor $i$ in layer $j$, i.e., $u^*_{i,j} = \max \{\tilde{u}_{i,j,1},\ldots,\tilde{u}_{i,j,K} \}$ (alternative definitions for $\rho_{j,j'}$ are possible; e.g., by considering the median instead of the maximum). This approach represents the network correlation after accounting jointly for social structure encoded within each layer, thanks to the MNLPM's hierarchical specification. Left panel in Figure \ref{fig_correlation_ic_wiring} shows credible intervals along with point estimates for all pairwise network correlations. We see that all pair of layers are positively correlated. This characteristic is particularly evident between Horseplay and Friendship (networks 1 and 3), which is consistent with the social dynamics described above.

Lastly, we test our correlation approach by considering a set of independent networks with no underlying structure. To do so, we independently generate $J=4$ Erd\"{o}s-R{\'e}nyi networks with $I=14$ actors and interaction probability 0.1, and then fit the MNLPM in order to obtain the pairwise network correlations. Right panel in Figure \ref{fig_correlation_ic_wiring} present the corresponding inference for these quantities. We see that the correlation between every pair of layers is negligible since the credible intervals are centered at zero, which is the expected behaviour for data with no correlation structure.

\subsubsection{Model fit}

\begin{figure}[!h]
	\centering
	\setlength{\tabcolsep}{0pt}
	\begin{tabular}{ccccc}
		& Horseplay & Arguments & Friendship & Antagonism \\
		\begin{sideways} \hspace{1.0cm} Raw data \end{sideways} &
		\includegraphics[scale = 0.38]{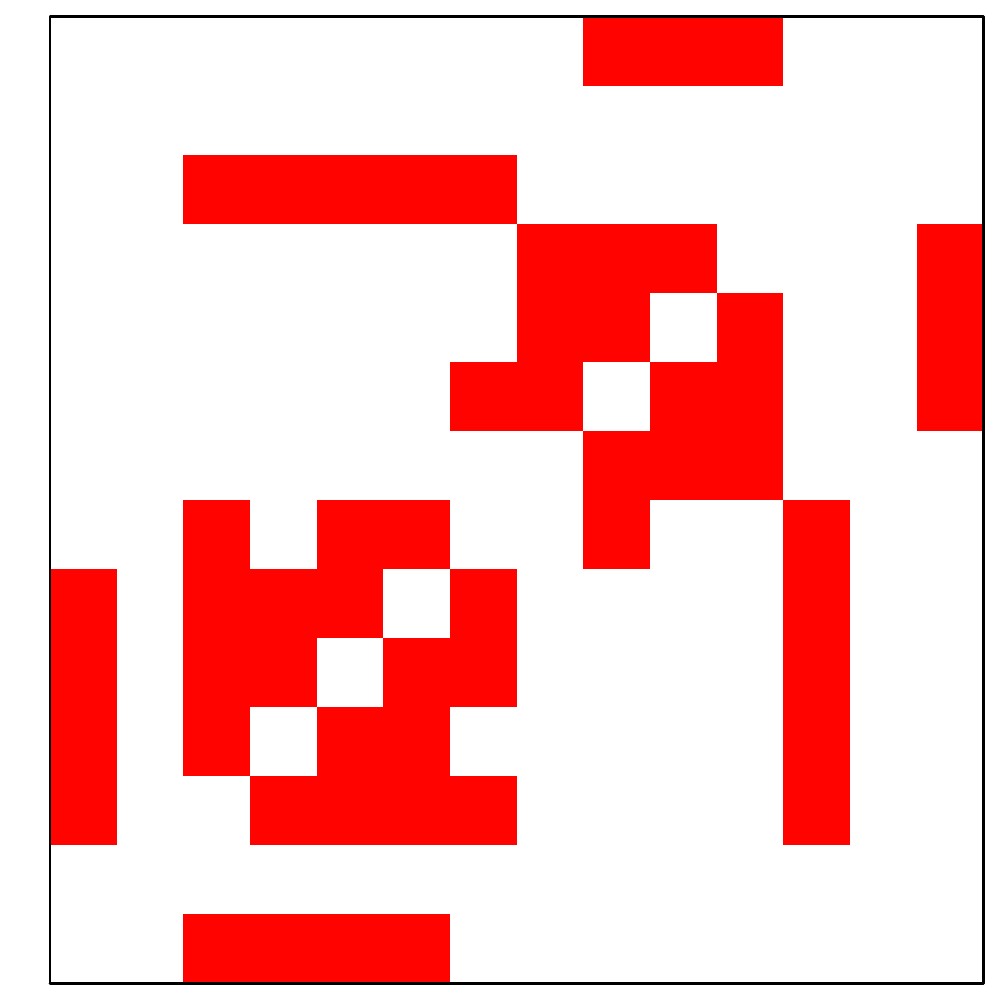} &
		\includegraphics[scale = 0.38]{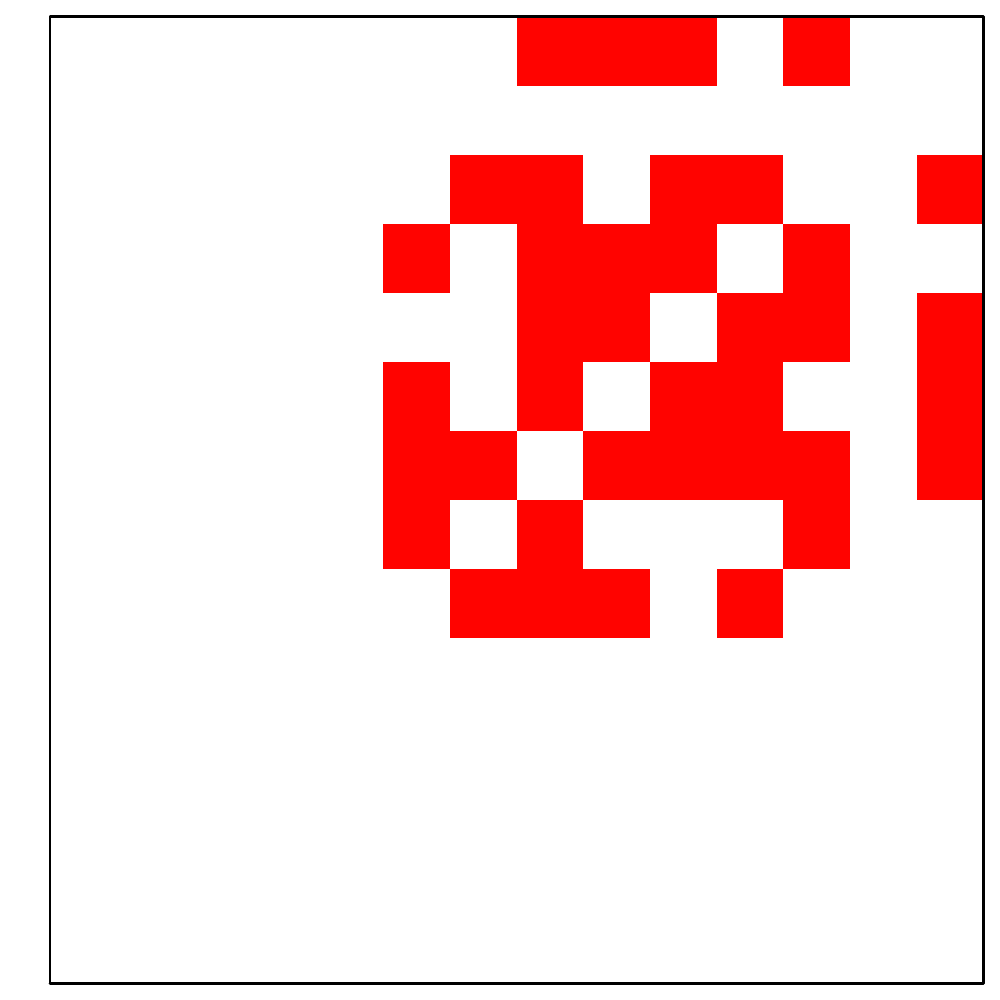} &
		\includegraphics[scale = 0.38]{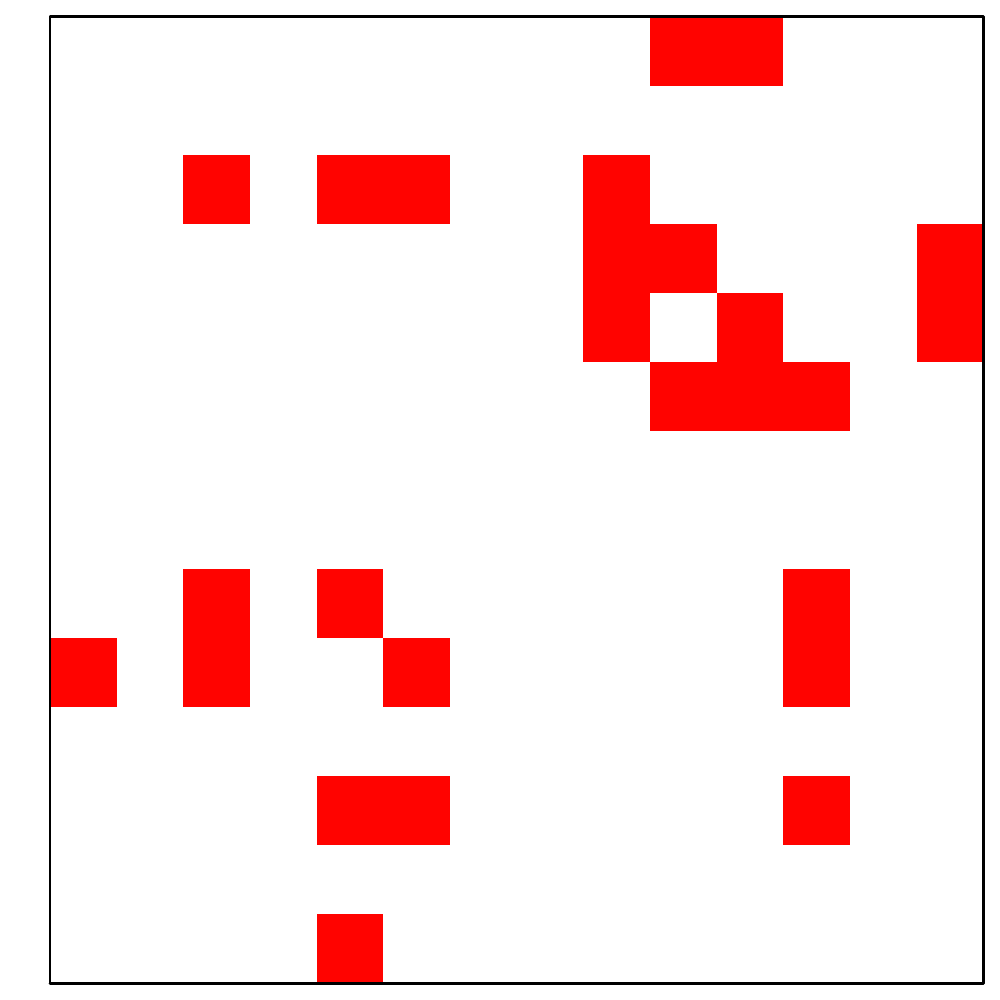} &
		\includegraphics[scale = 0.38]{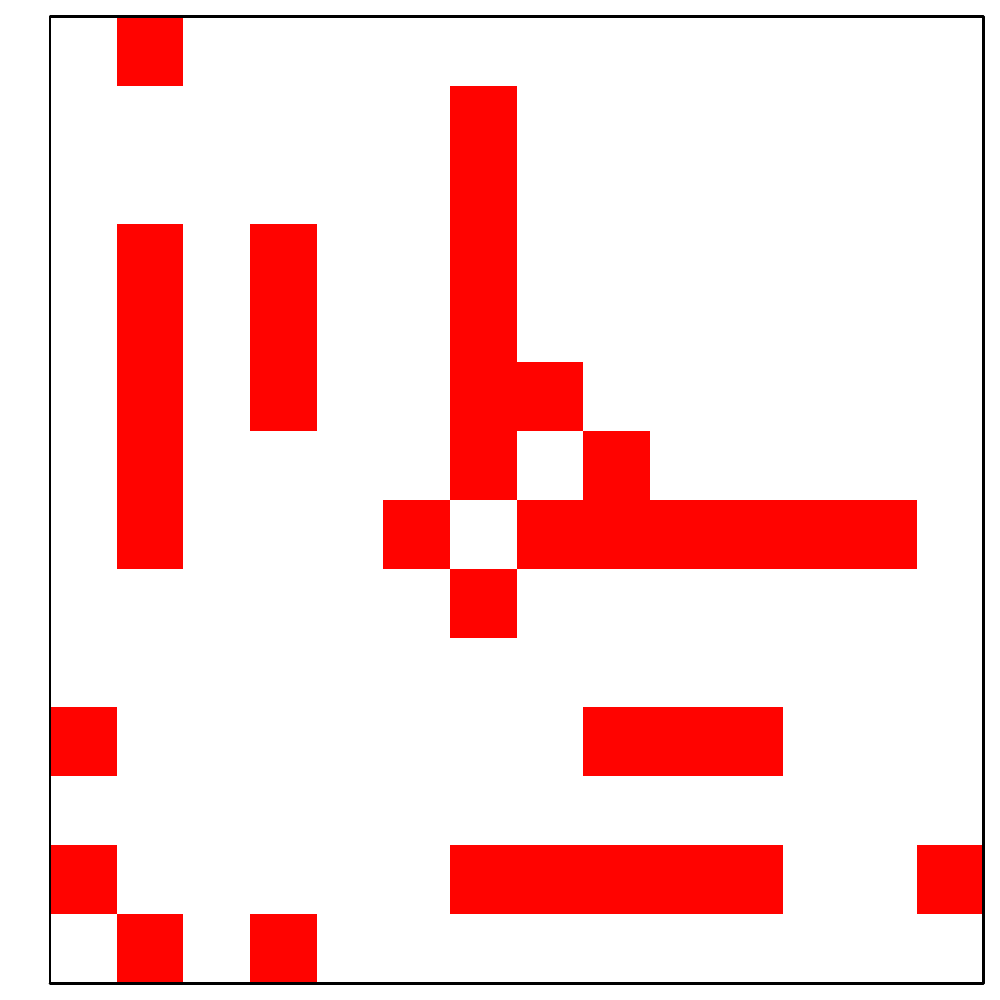} \\
		\begin{sideways} \hspace{0.8cm}  Probabilities \end{sideways}               &
		\includegraphics[scale = 0.38]{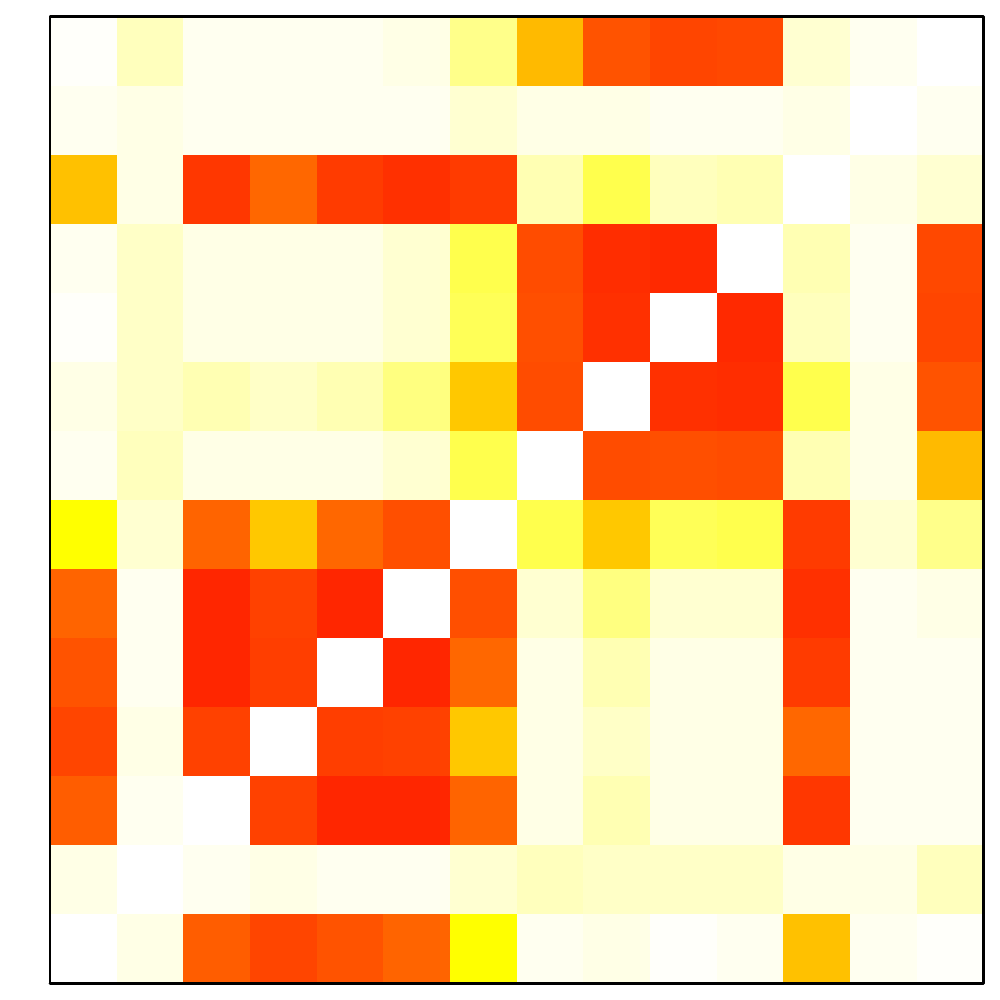}  &
		\includegraphics[scale = 0.38]{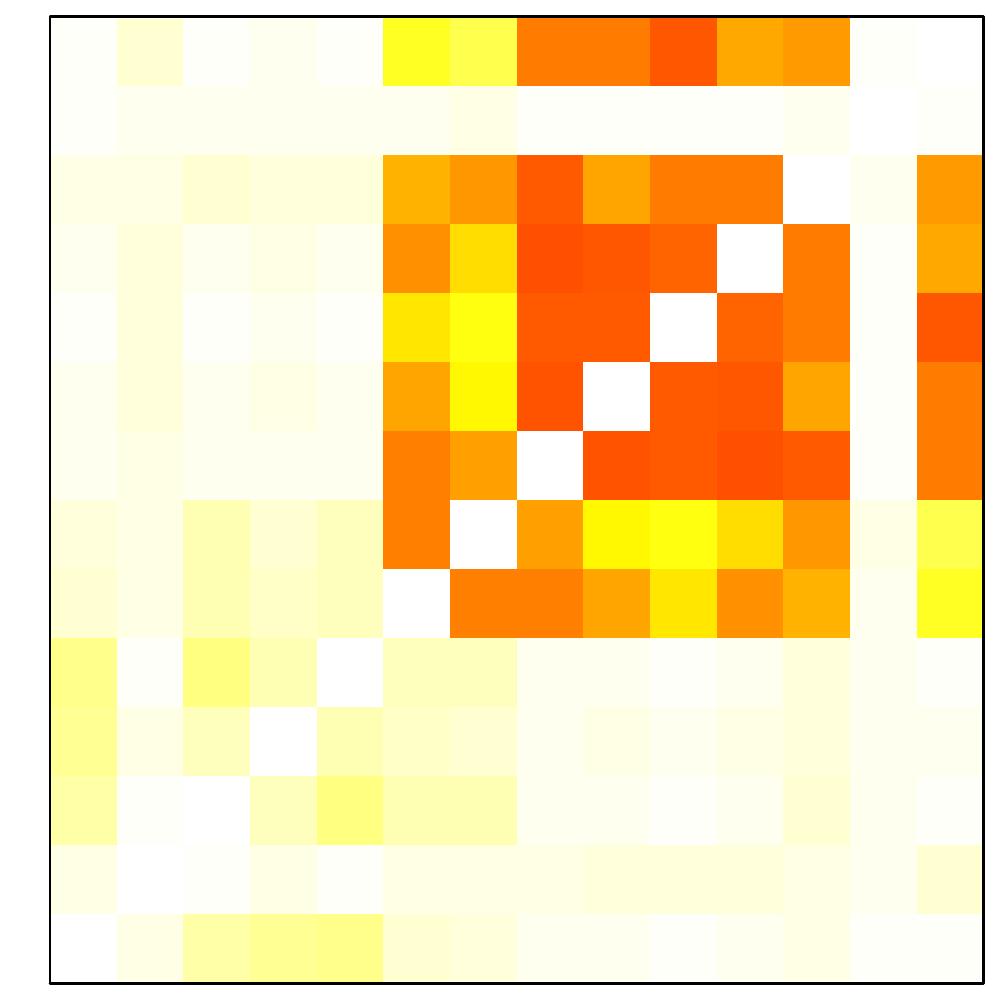}  &
		\includegraphics[scale = 0.38]{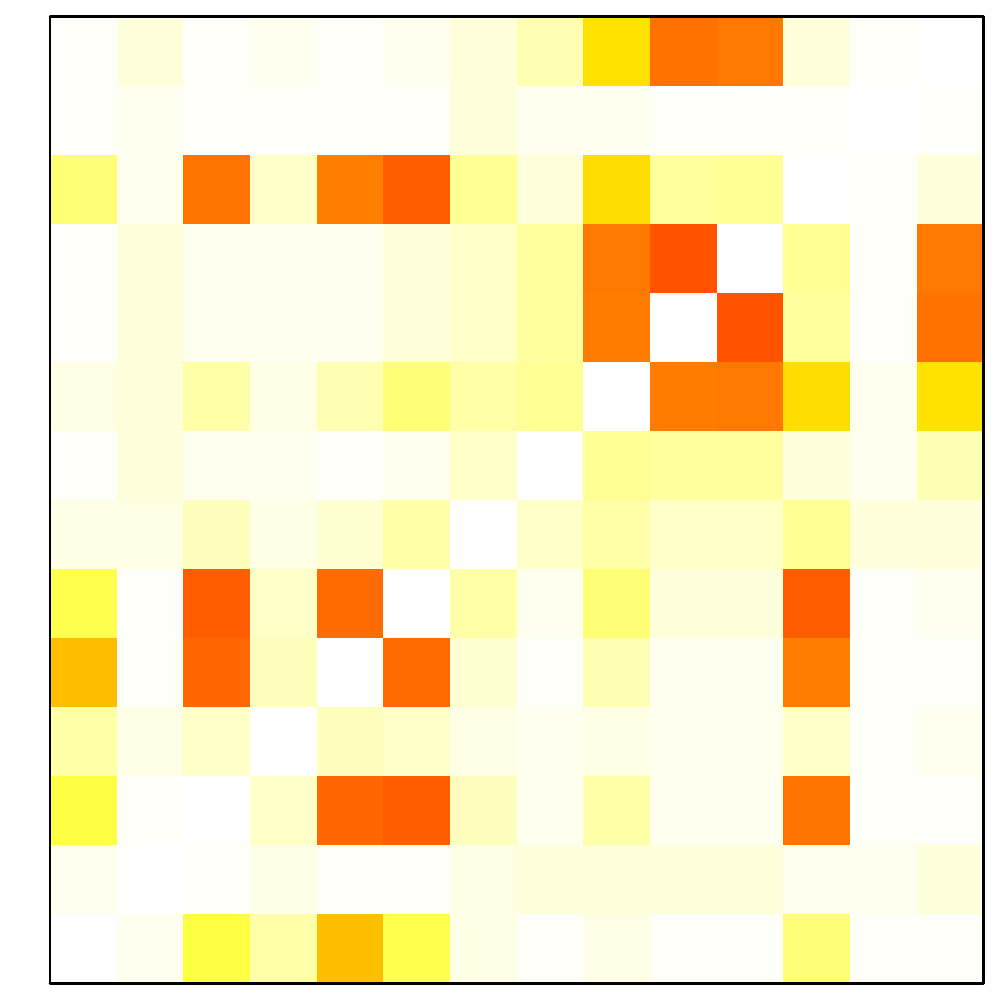}  &
		\includegraphics[scale = 0.38]{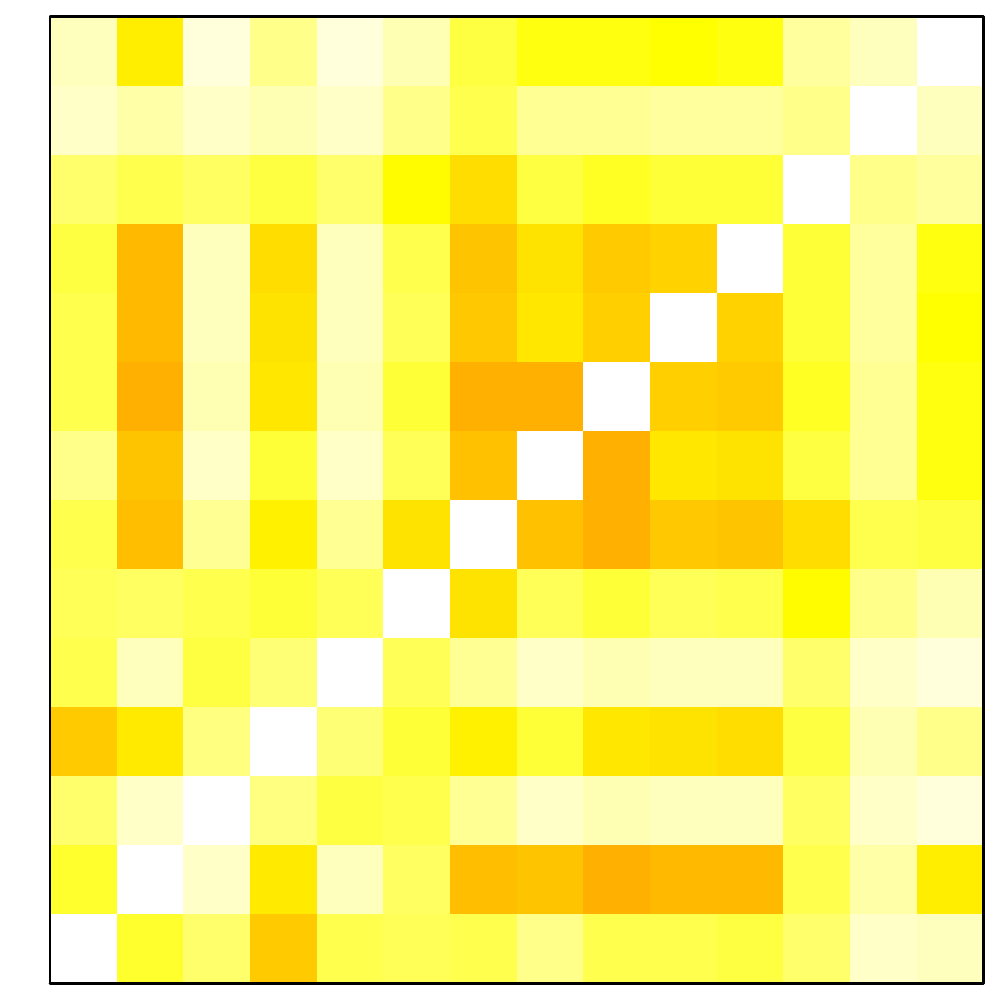}   \\
	\end{tabular}
	\includegraphics[scale=0.5]{./horizontal_bar}
	\caption{Multilayer network data $y_{i,i',j}$ and probability of interaction posterior means $\expec{\vartheta_{i,i',j}\mid\Y}$, for the bank wiring room data.}
	\label{fig_interaction_probabiliries}
\end{figure}

We assess our MNLPM fit using both in-sample and out-of-sample metrics. Our in-sample assessment relies on two approaches. First, we compare the observed data $y_{i,i',j}$ against the corresponding probability of interaction posterior means $\expec{\vartheta_{i,i',j}\mid\Y}$.
We see in Figure \ref{fig_interaction_probabiliries} that point estimates concur with the raw data, which clearly suggests that the model fits the data well in terms of reproducibility.

\begin{figure}[!t]
	\centering
	\subfigure[Density]               {\includegraphics[scale=0.4]{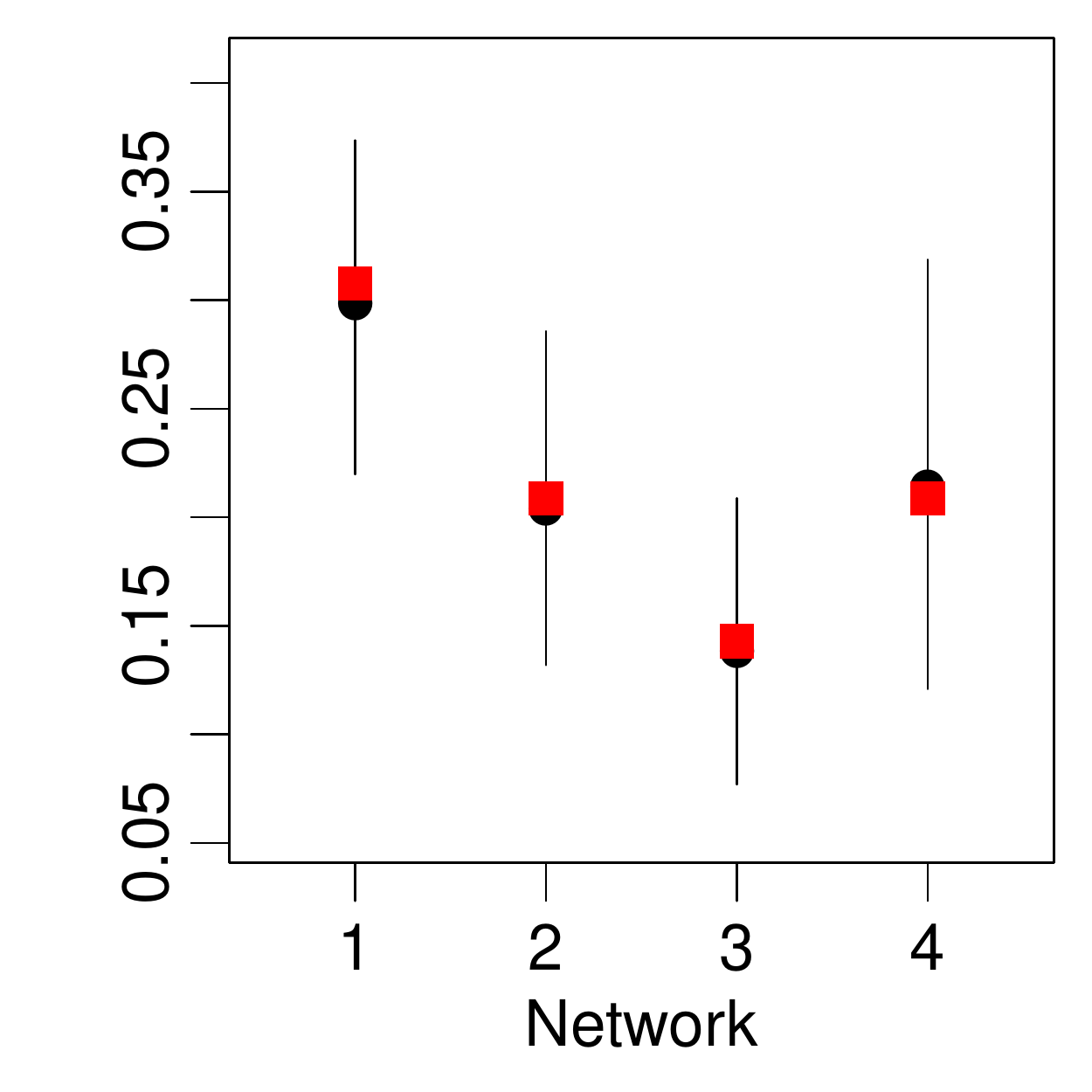}}
	\subfigure[Clustering coeff.]     {\includegraphics[scale=0.4]{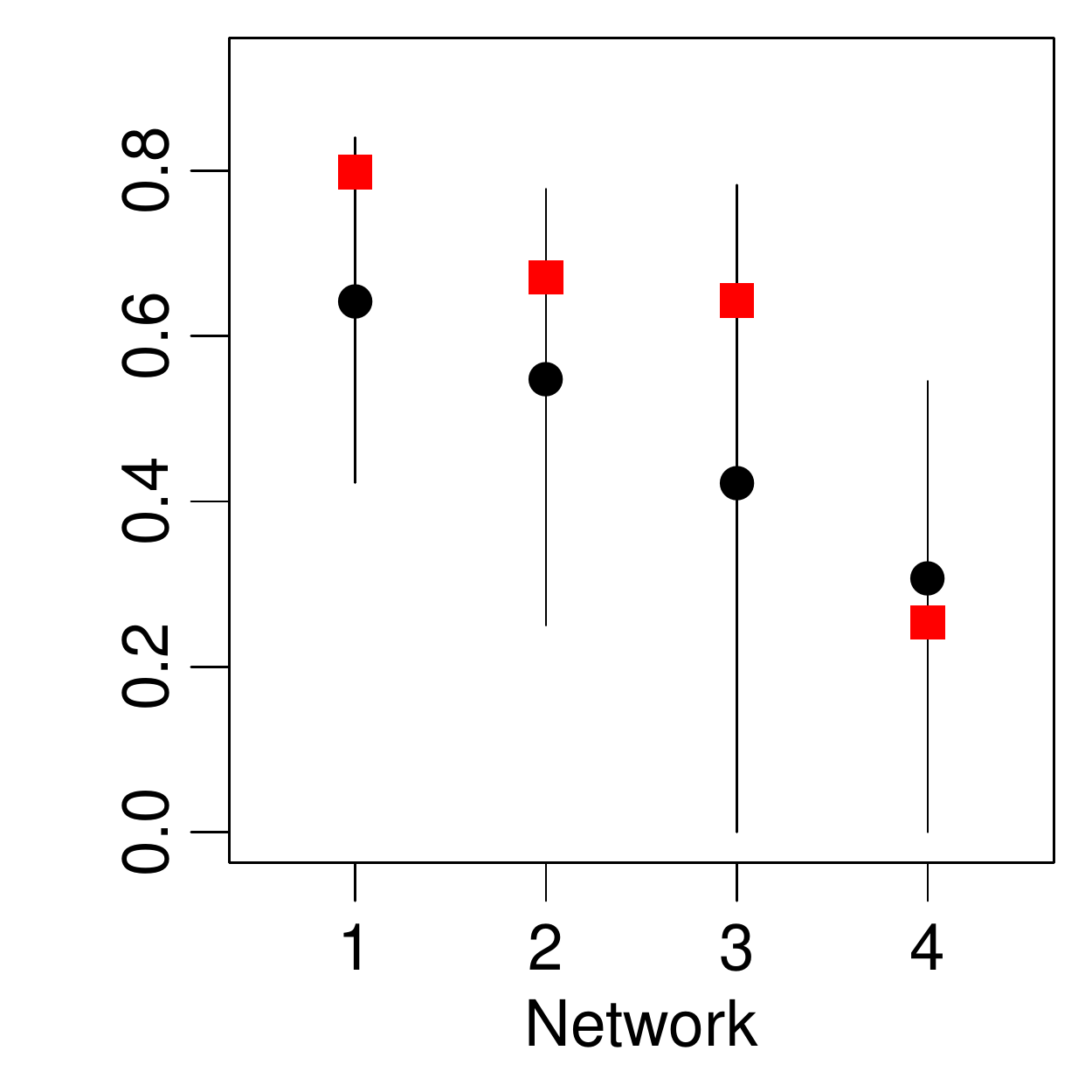}}
	\subfigure[Assortativity]         {\includegraphics[scale=0.4]{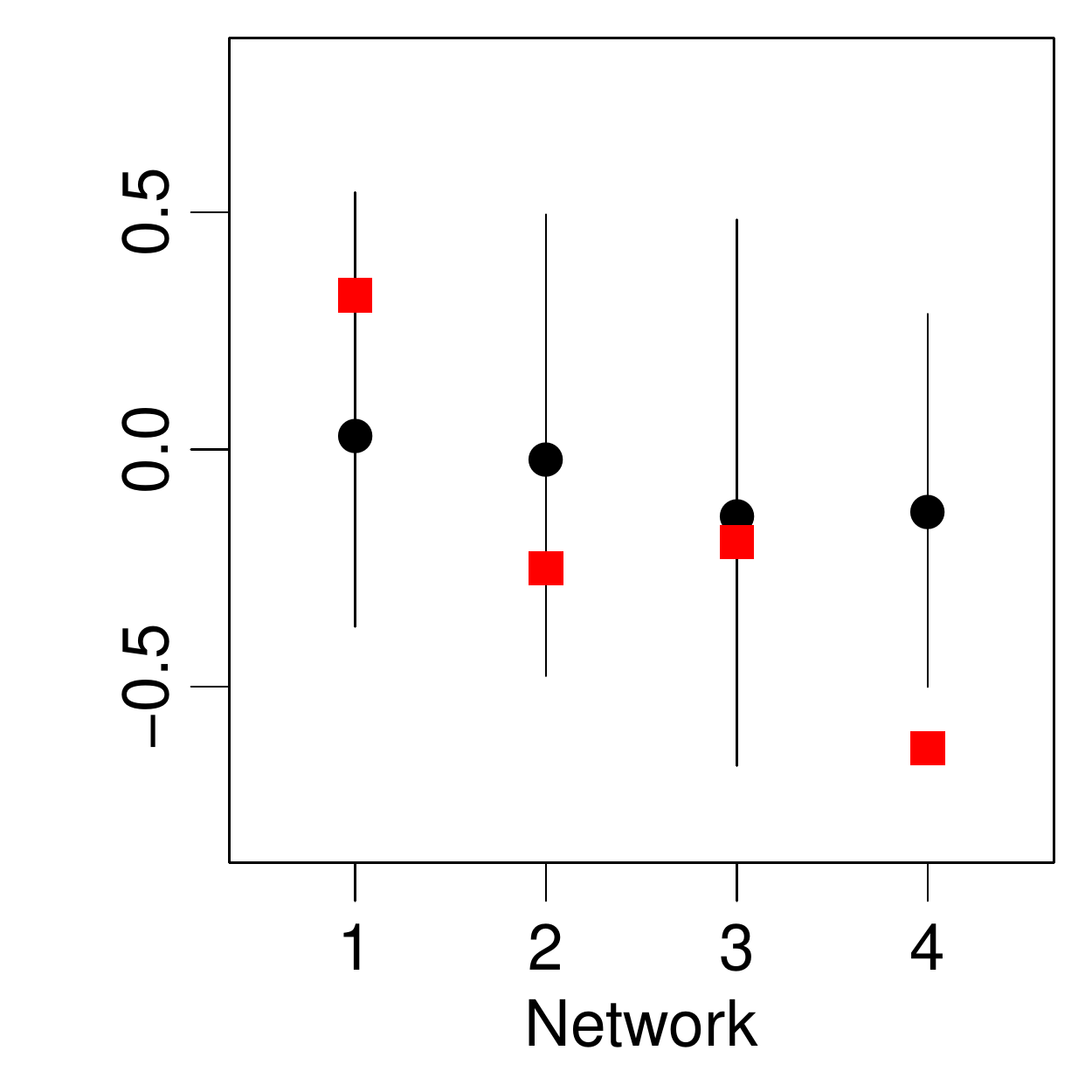}}
	\subfigure[Mean geodesic distance]{\includegraphics[scale=0.4]{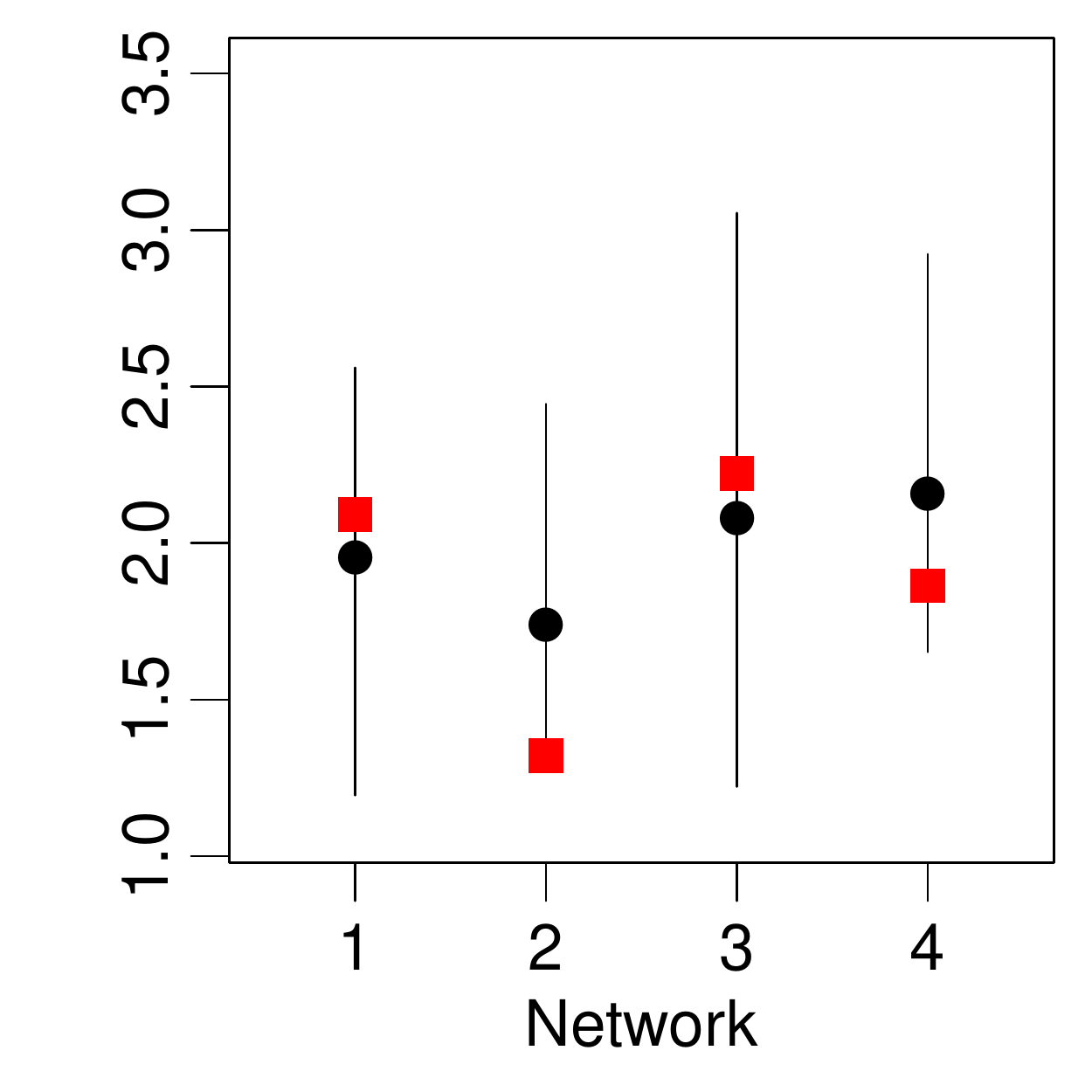}}
	\subfigure[Mean eigen-centrality] {\includegraphics[scale=0.4]{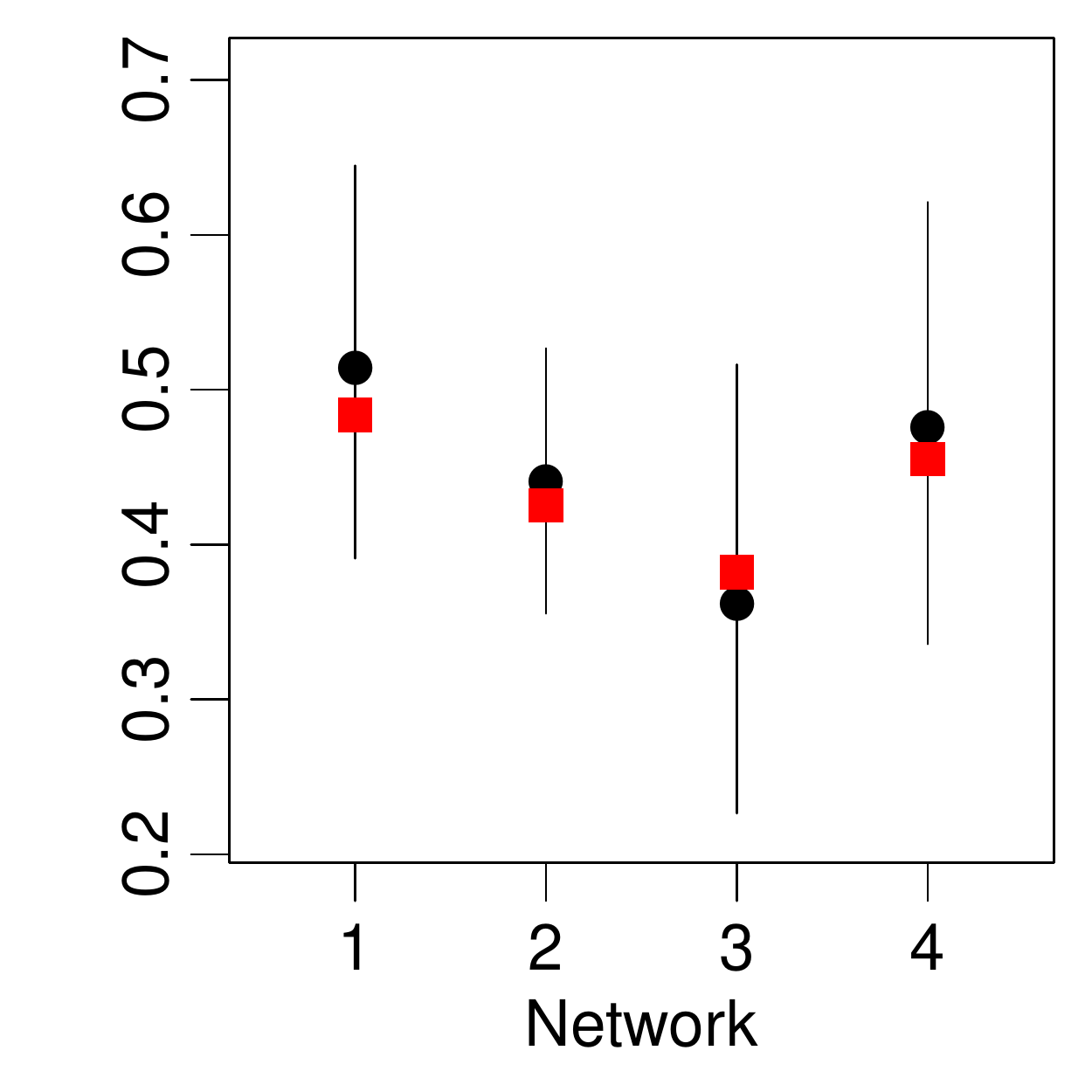}}
	\subfigure[Mean degree]           {\includegraphics[scale=0.4]{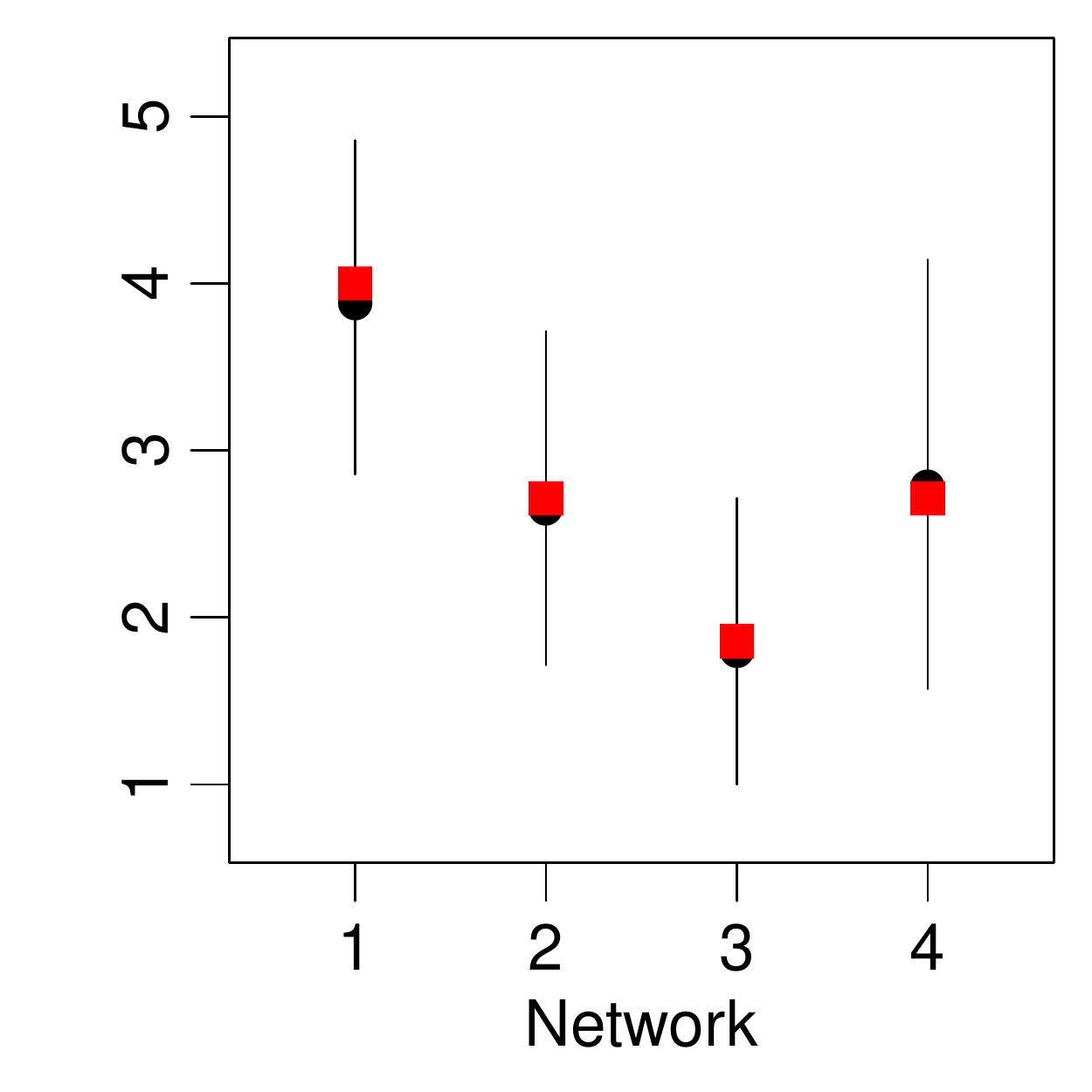}}
	\caption{95\% credible intervals, posterior means (black circle), and observed values (red square) associated with the empirical distribution of a battery of summary statistics, based on 10,000 replicas of the bank wiring room dataset (participation in horseplay, network 1; participation in arguments about open windows, network 2; friendship, network 3; and antagonistic behaviour, network 4).} \label{fig_goodness_wiring}
\end{figure}

Now, following \citet{gelman-2014}, we further explore the in-sample fit of each model by replicating pseudo-data from the fitted model and calculating a variety of summary statistics for each sample, whose distributions are then compared against their values in the original sample. Figure \ref{fig_goodness_wiring} shows credible intervals along with point estimates for a set of relevant network measures, including the density, assortativity, and clustering coefficient, among others (see \citealp{kolaczyk2014statistical} for details about these structural summaries). Note that the model appropriately captures these structural features (perhaps with the exception of the assortativity in Antagonism), since observed values belong to the corresponding credible intervals; even most of the estimates virtually coincide with the observed values. Thus, pseudo-data generation also provides evidence of proper in-sample properties in favour of our model. Finally, the out-of-sample predictive performance of the model is presented in Section \ref{sec_CV}.

\subsection{Friendship data}

In this section we develop a formal test to assess the level of ``agreement'' (as opposed to ``accuracy'', which requires the definition of an external ``gold standard''), between an actor's self perception of their own position in a social environment and that of other actors embedded in the same system (e.g., \citealp{swartz-2015}, \citealp{sewell2019latent}, \citealp{sosa2021cognitive}). Our approach relies on the hierarchical structure of the MNLPM, which allow us to define a measure of cognitive agreement. The posterior distribution of such a measure makes possible to identify those individuals whose position in the social space agrees with the judgements of other actors.

A cognitive social structure (CSS) is defined by a set of cognitive judgements that subjects form about the relationships among actors (themselves as well as others) who are embedded in a common environment. Hence, each subject reports a full description of the social network structure. We consider a CSS reported by \citet{krackhardt-1987} in which $I=21$ management personnel in a high-tech machine manufacturing organization were observed in order to evaluate the effects of a recent management intervention program. Each person was asked to fill out a questionnaire indicating not only who he/she believes his/hers friends are, but also his/her perception of others friendships. Thus, we have a collection composed of $J=21$ undirected binary networks $\mathbf{Y}_1, \ldots , \mathbf{Y}_I$, with $\mathbf{Y}_j=[y_{i,i',j}]$, defined over a common set of $I=21$ actors, such that $y_{i,i',j} = 1$ if $i$ and $i'$ are friends of each other, and $y_{i,i',j} = 0$ otherwise. Some attribute information about each executive was also available, including corporate level (president, vice-president, or general manager), and department membership (there are four departments labelled from 1 to 4; the CEO is not in any department). Such information can potentially be included in the analysis (see Section \ref{sec_discussion} for details).

\begin{figure}[!b]
	\centering
	\subfigure[Actor 7]  {\includegraphics[scale=0.15]{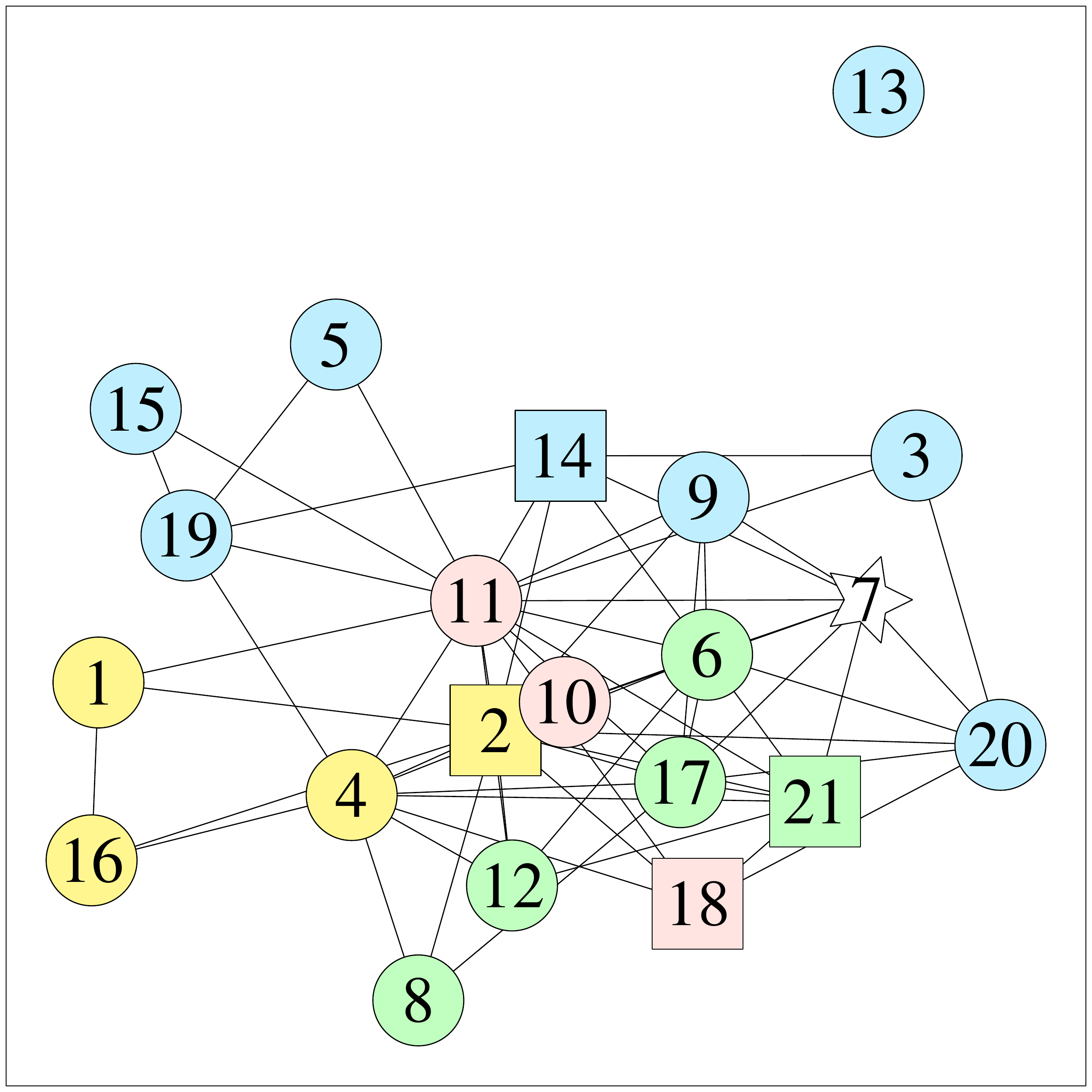}}
	\subfigure[Actor 14] {\includegraphics[scale=0.15]{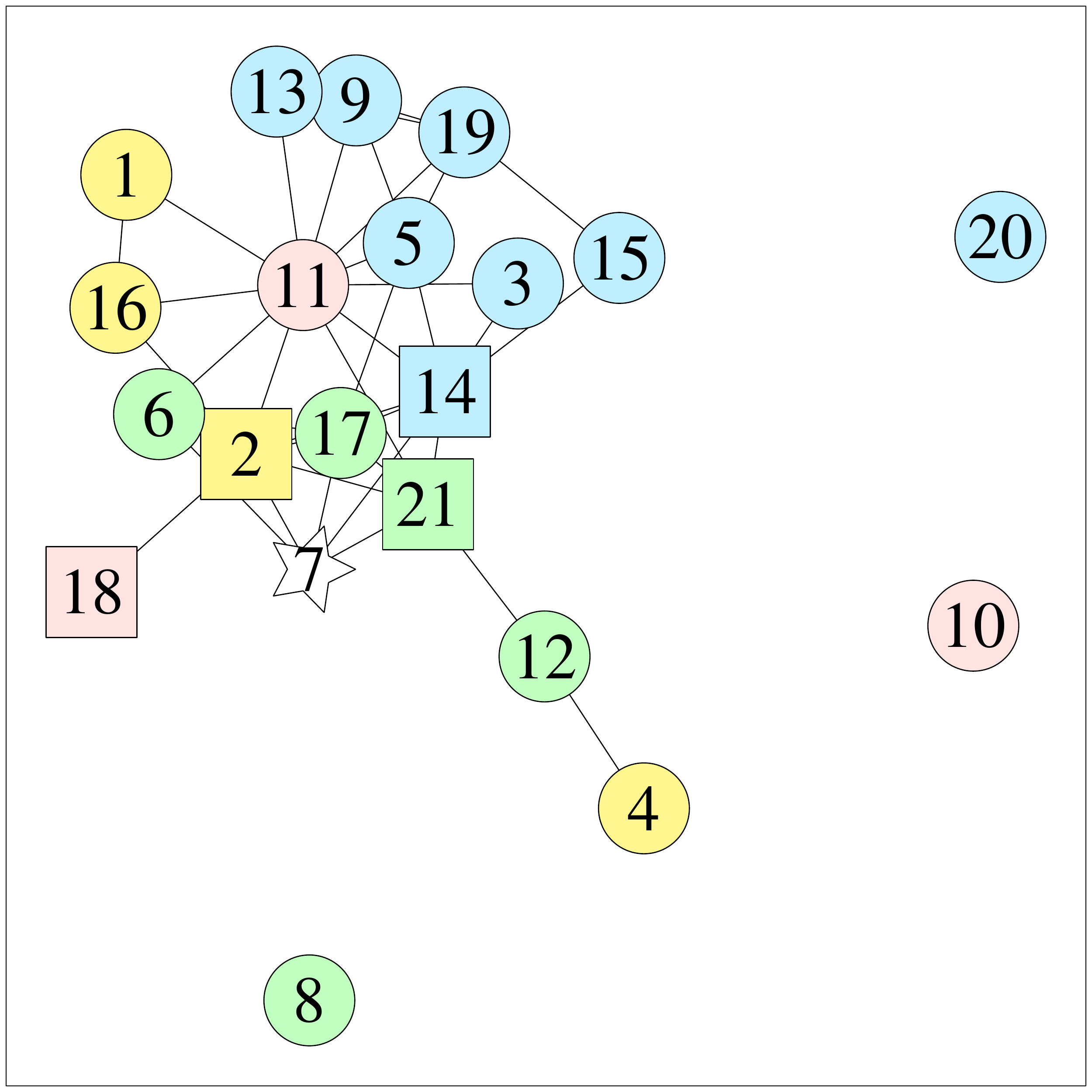}}
	\subfigure[Actor 17] {\includegraphics[scale=0.15]{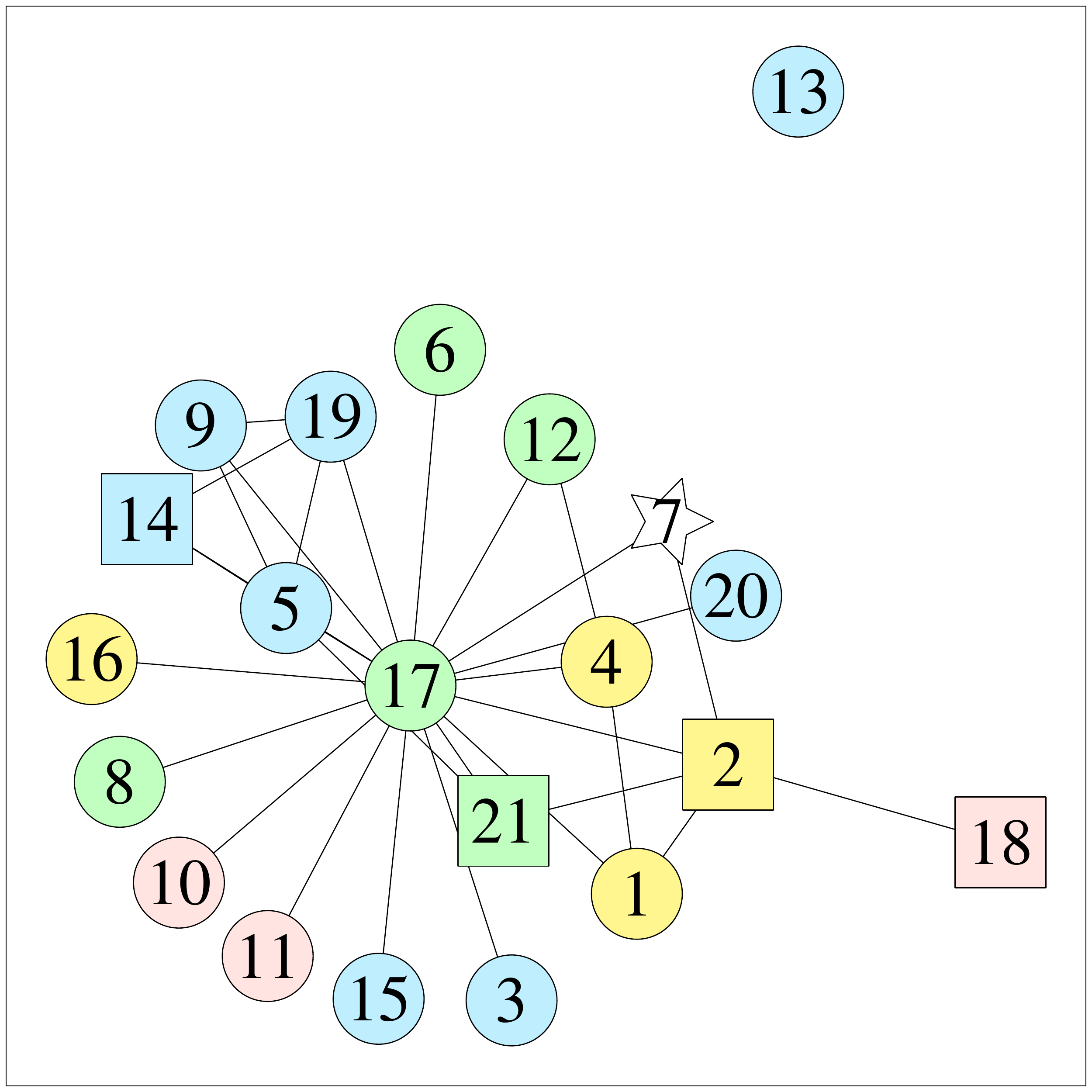}}
	\subfigure[Consensus] {\includegraphics[scale=0.15]{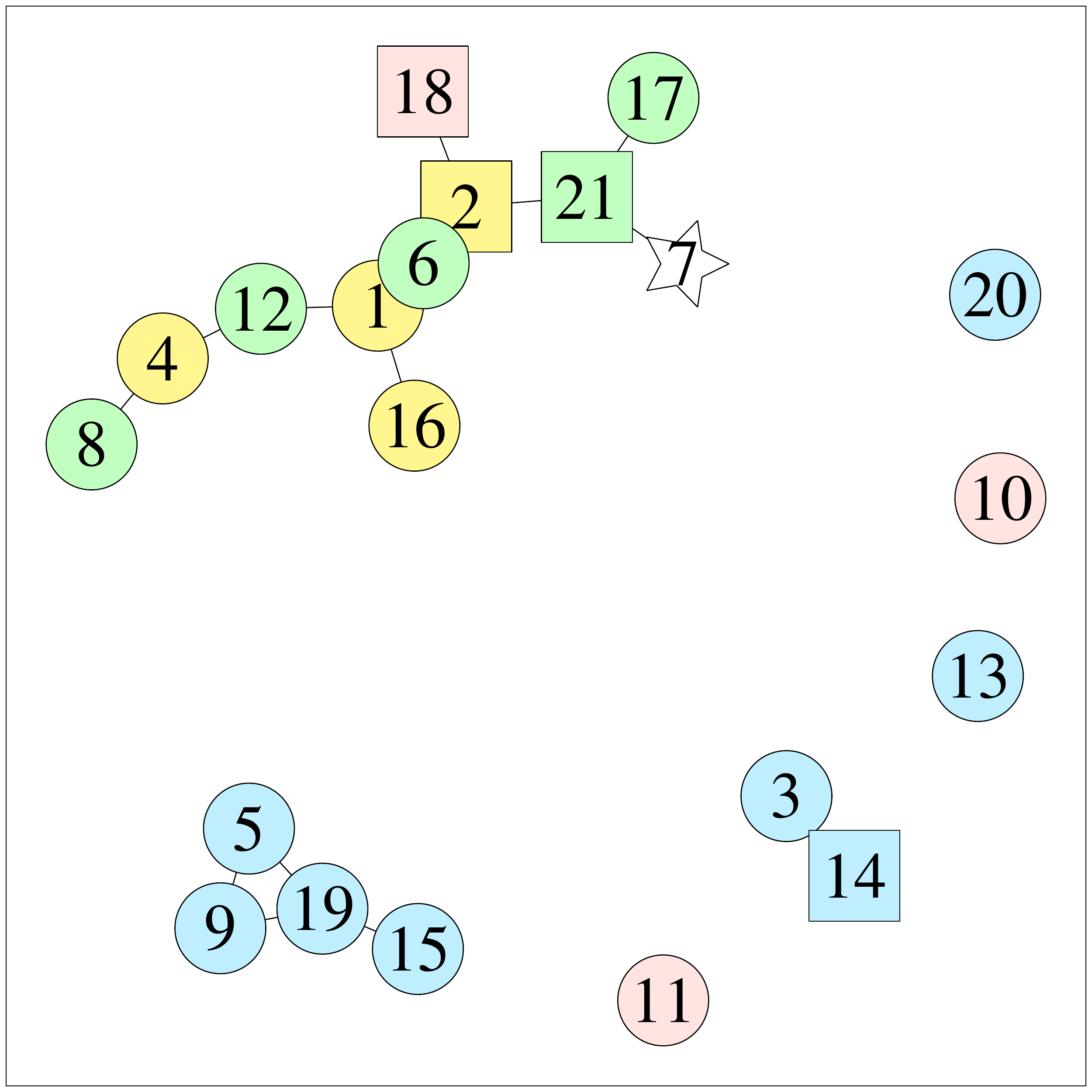}}
	\caption{Visualization of some networks in the friendship CSS data corresponding to actors 7, 14, and 17, along with the consensus network. Vertex shape indicates the executive's level in the company (star: president, actor 7; square: vice-presidents, actors 2, 14, 18, 21; and circles: managers), whereas vertex color indicates the executive's department in the company (the president does not belong to any department).}
	\label{fig_graphs_krackhardt21}
\end{figure}

Part of these multilayer network data along with the ``consensus'' network are represented in Figure \ref{fig_graphs_krackhardt21}. In this context, there is a link present between two actors according to the consensus if at least half of the personnel have reported that link. Note that even though the variability on the perceptions is not negligible, there are some commonalities across networks. For instance, more than half of the management personnel believes that actors 2 and 18 (both vice-presidents), actors 21 (vice-president) and 17 (manager) both in department 2, and actors 14 (vice-president) and 3 (manager) both in department 3, are friends. Furthermore, managers 3, 8, 9, and 20 only report less than six relations each, and manager 9, who only recognizes four friendships, is the only executive that considers himself with no friends. Senior executives (president and vice-presidents) report networks with more that 20 connections each.

Now, we describe the ``popularity'' of each executive in terms of how connected they are. Top panel in Figure \ref{fig_boxplots_krackhardt21} summarizes the degree distribution of actors across networks. In general, we see that all the executives perceive themselves as more popular than what they actually are according to the general opinion, with the exception of actors 8 and 9. On the other hand, actors 1, 2, 19, and 21 are perceived by the personnel as the executives with most connections. We also highlight the case of actors 10 and 17 who are highly egocentric in comparison with the consensus; in particular, actor 10 is perceived with no friends at all, but this actor believes just the opposite.

\subsubsection{Perception assessment}

\begin{figure}[!h]
	\centering
	\subfigure[Normalized degree]{\includegraphics[scale=0.6]{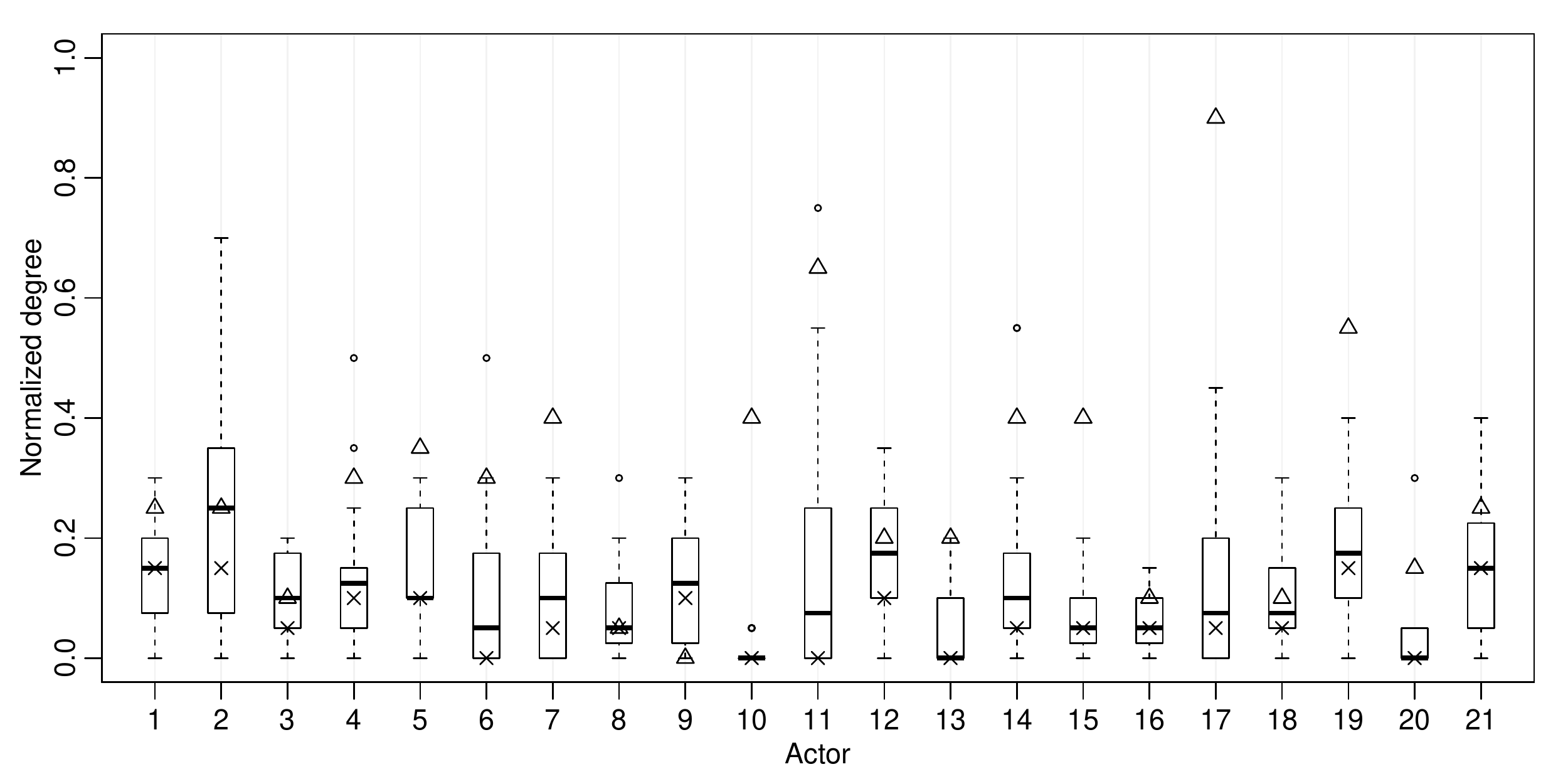}}
	\subfigure[Assessment index] {\includegraphics[scale=0.6]{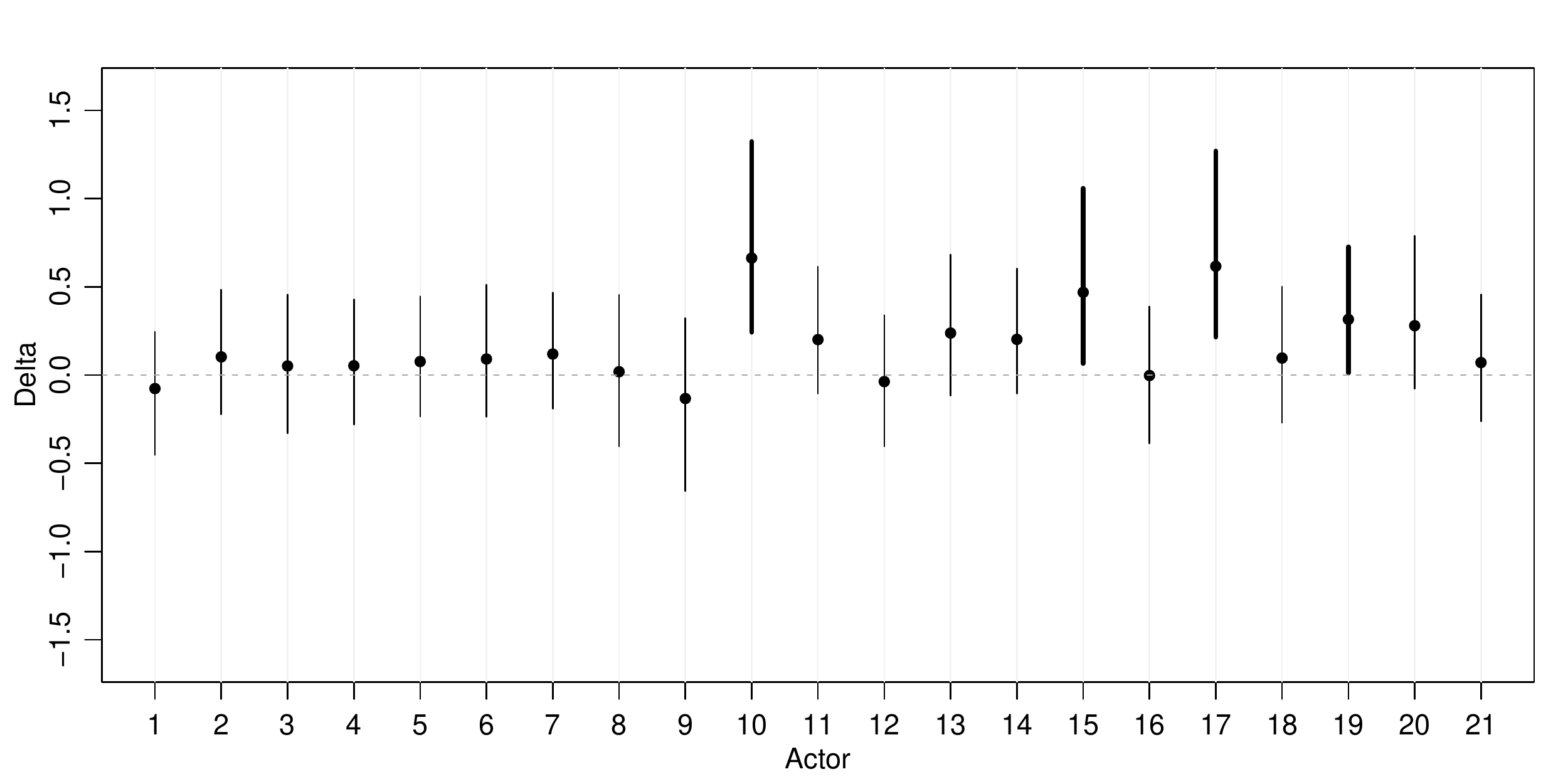}}
	\caption{Top panel: Normalized degree distribution across networks. The $i$-th boxplot summarizes the distribution of the degree for all reporters except $i$, while the self-perceived degree is represented by a triangle ($\bigtriangleup$) and the respective degree in the consensus network by a cross ($\times$). Bottom panel: 95\% credible intervals and posterior means for the distribution of the personal assessment parameters $\delta_i$. Thicker lines correspond to credible intervals that do not contain zero.}
	\label{fig_boxplots_krackhardt21}
\end{figure}

We consider the agreement question in which we ask whether an individual's perception of their relationships is the same as the perception that others hold. To answer this question, we define the assessment parameter $\delta_i$, for $i=1,\ldots,I$, as the difference between subject $i$'s self-assessment and the mean assessment of subject $i$ by others, i.e.,
$$\delta_i = \Big\|\, \tilde\uv_{i,i} \,\Big\| - \Big\|\, \frac{1}{I-1}\sum_{j\neq i}\tilde\uv_{i,j} \,\Big\|\,,$$
where $\tilde\uv_{i,j}$ is the Procrustes-transformed version of $\uv_{i,j}$. This quantity is an effort to parametrize the accuracy of self-assessment in perceiving ties.

Bottom panel in Figure \ref{fig_boxplots_krackhardt21} provides credible intervals along with point estimates for the personal assessment parameters $\delta_1,\ldots,\delta_I$, based on $B = 10,000$ samples of the posterior distribution obtained after thinning the original Markov chains every 10 observations and a burn-in period of 100,000 iterations, associated with the value of $K$ that optimizes the \textsf{WAIC} ($K = 6$).  We see that most actors have a slightly elevated view of themselves in terms of their capacity to befriend others, whereas very few have a negative view. On the other hand, actors 10, 15, 17, and 19 have a significant inflated perception of their ability to form friendship ties. Note that the results of this test are quite consistent with the exploratory data analysis discussed previously.

\begin{figure}[!h]
	\centering
	\subfigure[$\expec{\tilde\uv_{i,17}\mid\Y}$]{\includegraphics[scale=0.22]{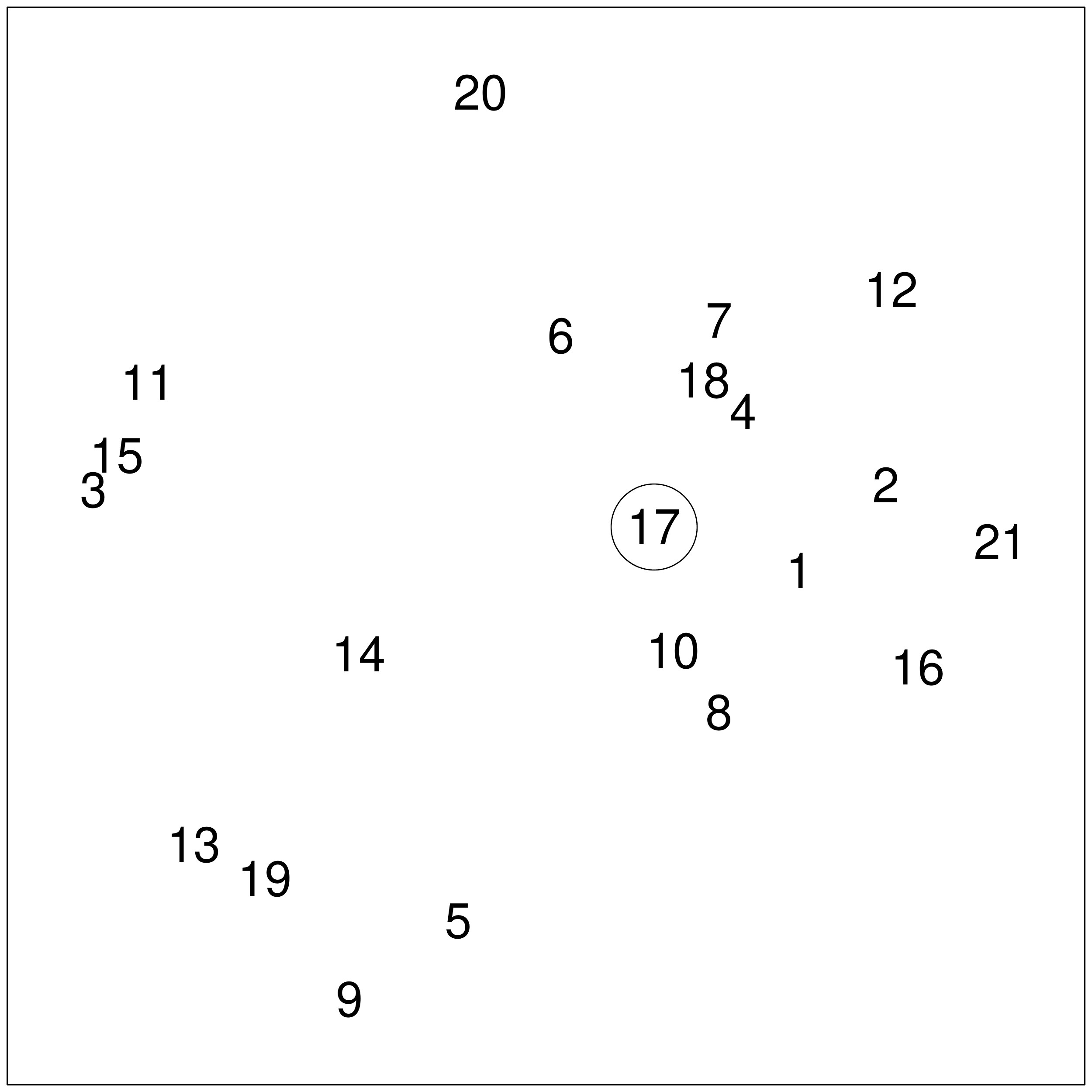}}
	\subfigure[$\expec{\tilde\uv_{17,i}\mid\Y}$]{\includegraphics[scale=0.22]{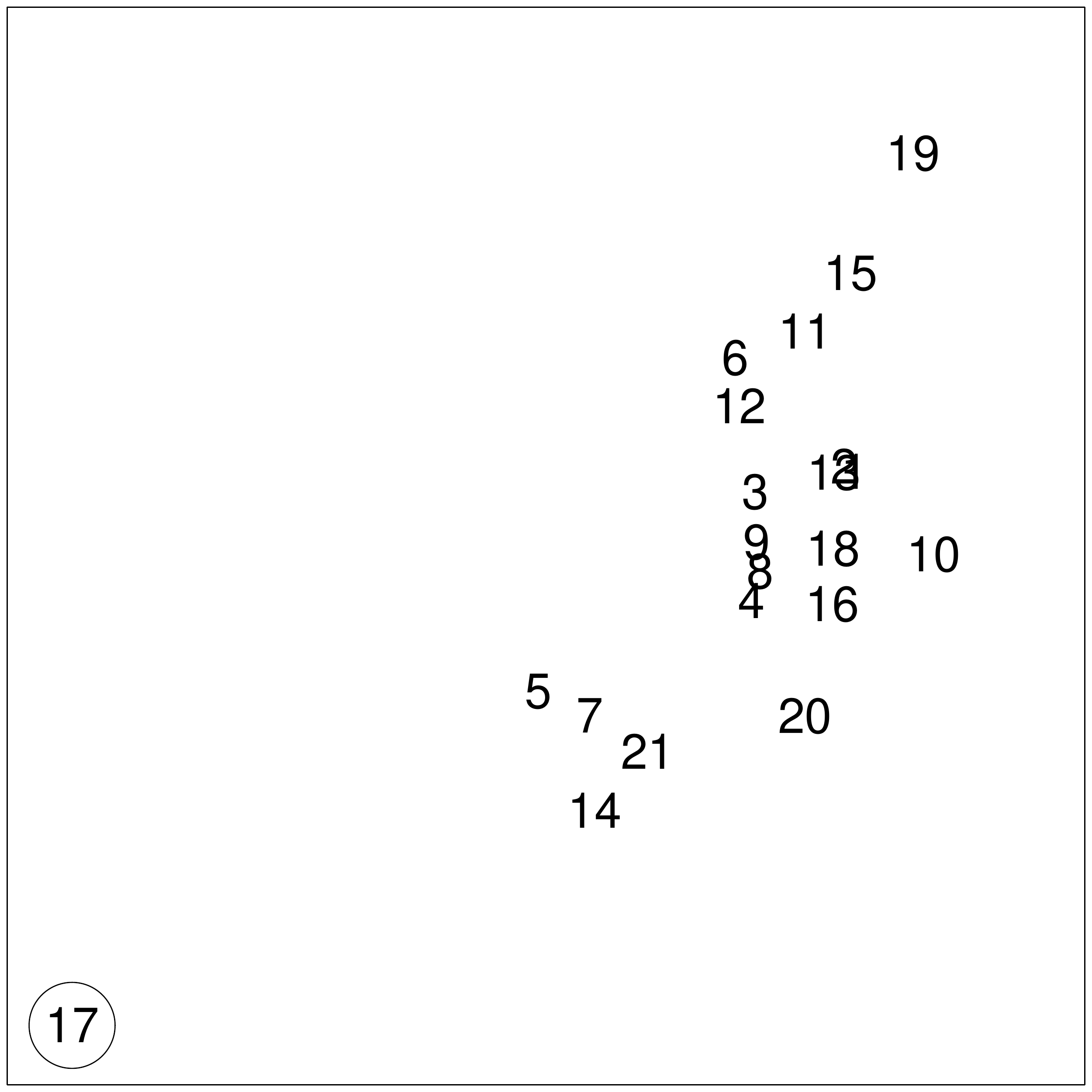}}
	\caption{Posterior means of procrustes-transformed latent positions along the two dimensions with highest variance for actor 17 (circled). Left panel: As perceived by actor 17, $\expec{\tilde\uv_{i,17}\mid\Y}$. Right panel: As perceived by all actors, $\expec{\tilde\uv_{17,i}\mid\Y}$.}
	\label{fig_projections_krackhardt21}
\end{figure}

Finally, in order to exemplify the social behaviour for an unskillful actor in perceiving relations, we show in Figure \ref{fig_projections_krackhardt21} Procrustes-transformed latent positions estimates along the two dimensions with highest variance for actor 17 (who clearly has a misleading view of his/hers surroundings according to the test), as perceived by this actor, $\expec{\tilde\uv_{i,17}\mid\Y}$, and as perceived by all actors (including him/herself), $\expec{\tilde\uv_{17,i}\mid\Y}$. These plots are consistent with those from Figure \ref{fig_boxplots_krackhardt21}. Actor 17 see him/herself in quite a ``central'' position of the friendship relations; however, according with the general opinion, actor 17 is clearly isolated from the others. This is again consistent with the test showed in Figure \ref{fig_boxplots_krackhardt21}.

\section{Prediction}\label{sec_CV}

As an additional goodness-of-fit assessment, we carry out cross-validation experiments on several multilayer network datasets (see Table \ref{tab_datasets}) exhibiting different kinds of actors, sizes, and relations. More specifically, we performed a five-fold cross-validation (CV) in which five randomly selected subsets of roughly equal size in the dataset are treated as missing and then predicted using the rest of the data.

\begin{table}[!h]
	\centering
	\begin{tabular}{llccc}  
		\hline
		Acronym & Reference & Actors & Layers & Edges  \\ 
		\hline
		\textsf{wiring} & \cite{roethlisberger2003management} &  14 & 4 & 79   \\
		\textsf{tech}  & \cite{krackhardt-1987}               &  21 & 21& 550  \\
		\textsf{seven}  & \cite{vickers1981representing}      &  29 & 3 & 222  \\
		\textsf{girls}  & \cite{steglich2006applying}         &  50 & 3 & 119  \\           		                
		\textsf{aarhus} & \cite{magnani2013combinatorial}     &  61 & 5 & 620  \\
 		\textsf{micro}  & \cite{banerjee2013diffusion}        &  77 & 6 & 903  \\		
		\hline
	\end{tabular}\caption{Multilayer network datasets for which a series of cross-validation experiments are performed using independently fitted LPMs (baseline) and our MNLPM. Note that \textsf{wiring} and \textsf{tech} are widely analysed in Section \ref{sec_ilustrations}. Also, \textsf{micro} corresponds to the network data of village number 10.}\label{tab_datasets} 
\end{table}

\begin{table}[!b]
	\centering
	\begin{tabular}{l|ccc|ccc}  
		\hline
		Measure& \multicolumn{3}{c|}{\textsf{AUC}} & \multicolumn{3}{c}{\textsf{WAIC}} \\
		\hline\hline
		Acronym & IFLPM & GMLPM & MNLPM & IFLPM & GMLPM & MNLPM \\ 
		\hline
		\textsf{wiring} & 0.905 & 0.823 & 0.895 & 201.2  & 293.8  & 197.1  \\
		\textsf{tech}   & 0.804 & 0.882 & 0.904 & 2140.1 & 2318.8 & 1492.2 \\
		\textsf{seven}  & 0.888 & 0.957 & 0.958 & 470.7  & 490.4  & 345.5  \\
		\textsf{girls}  & 0.781 & 0.879 & 0.882 & 352.8  & 525.2  & 310.6  \\
		\textsf{aarhus} & 0.922 & 0.926 & 0.949 & 1695.8 & 2611.1 & 1513.0 \\
		\textsf{micro}  & 0.773 & 0.936 & 0.932 & 4939.8 & 3799.9 & 3566.5 \\		
		\hline
	\end{tabular}
	\caption{Average \textsf{AUC}s and \textsf{WAIC} corresponding to the prediction of missing links in a series of CV experiments to assess the predictive performance of IFLPM, GMLPM, and MNLPM, using each dataset provided in Table \ref{tab_datasets}.}\label{tab_cv} 
\end{table}

We summarize our findings in Table \ref{tab_cv}, where we report the average area under the receiver operating characteristic curve (\textsf{AUC}) and the \textsf{WAIC} for each dataset described in Table \ref{tab_datasets}. The values correspond to the prediction of missing links using independently fitted LPMs (IFLPM), our MNLPM, and also, a variant of MNLPM that is very reminiscent of \cite{gollini2016joint}. The latter, referred to as GMLPM, considers unique latent positions with no hierarchical structure in such a way that $\vartheta_{i,i',j} = \Phi(\zeta_j - e^{\theta_j}\|\,\uv_i - \uv_{i'}\,\|)$. In this context, the \textsf{AUC} is a measure of how well a given model is capable of predicting missing links (higher \textsf{AUC} values are better). We report the \textsf{AUC} for the models with the optimal value of $K$ according to the \textsf{WAIC} criteria. As before, our predictions are based on $B = 10,000$ samples of the posterior distribution obtained after thinning the original Markov chains every 10 observations and a burn-in period of 100,000 iterations.

We see that MNLPM is clearly the best alternative in terms of both prediction and goodness-of-fit. Specifically, the out-of-sample performance of IFLPM and NMLPM is practically the same for \textsf{wiring}, as well as that of GMLPM and MNLPM for \textsf{seven}, \textsf{girls}, and \textsf{micro}. For all the other datasets, MNLPM has a better predictive behaviour than its competitors. Such an effect is particularly clear when fitting MNLPM as opposed to IFLPM, which provides even more evidence about why considering our hierarchical prior as in MNLPM is beneficial. On the other hand, not surprisingly, the \textsf{WAIC} of GMLPM is the highest for every dataset probably because this model forces the latent positions to be the same across layers, which is not the case for both IFLPM and MNLPM. Either way, MNLPM, once again, shows better information criterion values at all instances. These results strongly suggest that the predictive potential as well as the inner flexibility of MNLPM are indeed comparable with or even better than those offered by alternative approaches.

\section{Discussion}\label{sec_discussion}

This paper presents a novel approach to modeling multilayer network data with a method that encourages the flow of information across networks, as opposed to an independent characterization of each of them. Our proposal is based on a natural hierarchical extension of a latent space distance model, which provides a direct description of actors' roles within and across networks at global and specific levels. Furthermore, our experiments provide sufficient empirical evidence to establish that our approach is highly competitive in terms of prediction and goodness-of-fit.

Our MNLPM is susceptible to many generalizations. First, the model can be extended to represent patterns in the data related to known covariates by letting $\vartheta_{i,i',j} = \Phi( \xv_{i,i'}^\trans\zev_j - e^{\theta_j}\|\, \uv_{i,j} -  \uv_{i',j} \|\, )$, where $\xv_{i,i'} = (x_{i,i',1},\ldots,x_{i,i',P})$, in addition to a global intercept, is a vector of predictors that incorporates known attributes associated with actors $i$ and $i'$, and $\zev_j = (\zeta_{j,1},\ldots,\zeta_{j,P})$ is an unknown vector of fixed effects. Furthermore, in order to represent more general combinations of structural equivalence and homophily in varying degrees, it is also possible to consider other types of latent effects as in a factorial model, by letting $\vartheta_{i,i',j} = \Phi(\zeta_j + \uv_{i}\LAM_j\uv_{i'})$, where $\LAM_j = \textsf{diag}(\lambda_{j,1},\ldots,\lambda_{j,K})$ is a $K\times K$ diagonal matrix. Lastly, the model can also be modified to handle undirected networks by distinguishing latent ``sender'' positions, $\uv_{i,j}$, and latent ``receiver'' positions, $\vv_{i,j}$, which leads to $\vartheta_{i,i',j} = \Phi(\zeta_j - e^{\theta_j}\|\,\uv_{i,j} - \vv_{i',j} \|\,)$.

In the same spirit of \cite{green2009reversible}, we can also conceive a trans-dimensional version of the model that treats the latent dimension $K$ as a model parameter (as opposed to a fixed pre-specified quantity), which is quite challenging aside from the computational complexity, since in a varying-dimension case it is not clear how to provide a meaningful interpretation to the latent dimensions. Moreover, following \cite{guhaniyogi2018joint}, a truncation of a non-parametric process can be also incorporated into the model, but based on the authors' experience, results are likely to be quite similar.

Also, note that social positions might exhibit clustering patterns, which can be modelled directly by considering cluster assignment parameters $\xi_1,\ldots,\xi_I$ into the hierarchical specification of the model through a Categorical-Dirichlet prior (e.g., \citealp{handcock-2007}, \citealp{krivitsky-2008}, \citealp{krivitsky-2009}). Specifically, we can assume that all the actors in the system are clustered into $H$ groups, each of which occupies a position $\boldsymbol{\varphi}_h$ in the social space, $h = 1\ldots,H$. In this way, we can think of actor $i$'s average position $\etav_i$ as a Normal deviation from the group position to which it belongs, i.e., $\etav_i \mid \{\boldsymbol{\varphi}_h\},\kappa^2, \xi_i  \simind \Nor_K\left(\boldsymbol{\varphi}_{\xi_i}, \kappa^2\I\right)$, where $\xi_i=h$ means that actor $i$ belongs to cluster $h$. Nonparametric Bayes approaches in the same spirit of \cite{rodriguez2015bayesian} and \cite{d2019latent} are also possible.

Finally, we recommend consider alternative inference methods in order to account for ``big networks'', which is currently an active research area in computational statistics (e.g, \cite{gollini2016joint}, \citealp{ma2017exploration}, \citealp{spencer2020faster}, \citealp{aliverti2020stratified}).

\bibliography{references}
\bibliographystyle{apalike}

\appendix

\section{MCMC algorithm}\label{app_mcmc}

The posterior distribution is given by:
\begin{align*}
p(\UPS\mid\Y) &= \prod_{j, i<i'} \Ber\left(y_{i,i',j}\mid\Phi\left(\zeta_j - e^{\theta_j}\|\,\uv_{i,j} - \uv_{i',j}\|\,\right)\right) 
\times \prod_{i,j} \Nor_K\left(\uv_{i,j}\mid\etav_i, \sig^2\I\right) \\
&\times \prod_j \Nor_1\left(\theta_j\mid\mu_\theta, \tau^2_\theta\right) 
\times \prod_j \Nor_1\left(\zeta_j\mid\mu_\zeta, \tau^2_\zeta\right) 
\times \Nor_K\left(\etav_i\mid\nuv, \kappa^2\I\right)
\times \IGamd\left(\sig^2\mid a_\sig, b_\sig \right) \\
&\times \Nor_1\left(\mu_\theta\mid m_\theta, v^2_\theta\right) 
\times \IGamd\left(\tau^2_\theta\mid a_\theta, b_\theta \right)
\times \Nor_1\left(\mu_\zeta\mid m_\zeta, v^2_\zeta\right) 
\times \IGamd\left(\tau^2_\zeta\mid a_\zeta, b_\zeta \right) \\[2.5ex]
&\times \Nor_1\left(\nuv\mid \mv_{\nuv}, \V_{\nuv}\right) 
\times \IGamd\left(\kappa^2\mid a_\kappa, b_\kappa \right)\,.
\end{align*}

For a given set of fixed hyperparameters, $a_\sig, b_\sig, a_\zeta, b_\zeta, a_\theta,b_\theta, a_\kappa, b_\kappa, m_\zeta, v_\zeta, m_\theta, v_\theta, \mv_{\nuv}, \V_{\nuv}$, the algorithm proceeds by generating a new state $\UPS^{(b+1)}$ from a current state $\UPS^{(b)}$, $b = 1,\ldots,B$, as follows:
\begin{enumerate}
	
	\item Sample $\uv_{i,j}^{(b + 1)}$, $i = 1, \ldots, I$, $j = 1,\ldots,J$, according to an adaptive Metropolis-Hastings algorithm with the full conditional distribution:
	\begin{align*}
	p(\uv_{i,j}\mid\text{rest}) &\propto \prod_{i':i <i'}\Ber\left(y_{i,i',j}\mid\Phi(\zeta_j - e^{\theta_j}\|\,\uv_{i,j} - \uv_{i',j}\,\|)\right) \\ 
	&\times \prod_{i':i > i'}\Ber\left(y_{i',i,j}\mid\Phi(\zeta_j - e^{\theta_j}\|\,\uv_{i',j} - \uv_{i,j}\,\|)\right) \\
	&\times \Nor_K(\uv_{i,j}\mid\etav_i,\sig^2\I)\,.
	\end{align*}
	
	\item Sample $\etav_i^{(b+1)}$, $i=1,\ldots,I$, from:
	$$
	\etav_i\mid\text{rest} \sim \Nor_K\left(\left[\frac{1}{\kap^2} + \frac{J}{\sig^2}\right]^{-1} \left[ \frac{1}{\kap^2}\nuv  + \frac{1}{\sig^2}\sum_j\uv_{i,j} \right] , \left[\frac{1}{\kap^2} + \frac{J}{\sig^2}\right]^{-1}\I \right)\,.
	$$
	
	\item Sample $(\sigma^2)^{(b+1)}$ from:
	$$
	\sig^2\mid\text{rest} \sim \IGamd\left(a_\sig + \frac{I\,J\,K}{2}, b_\sig + \frac12 \sum_{i,j} (\uv_{i,j}-\etav_i)^\trans(\uv_{i,j}-\etav_i) \right)\,.
	$$
		
	\item Sample $\nuv^{(b+1)}$ from:
	$$\nuv\mid\text{rest} \sim \Nor_K\left( \left[\V_{\nuv}^{-1} + \frac{I}{\kappa^2}\I\right]^{-1}\left[\V_{\nuv}^{-1}\mv_{\nuv} + \frac{1}{\kappa^2}\sum_{i} \etav_i\right] , \left[\V_{\nuv}^{-1} + \frac{I}{\kappa^2}\I\right]^{-1} \right)\,.
	$$
	
	\item Sample $(\kappa^2)^{(b+1)}$ from:
	$$
	\kappa^2\mid\text{rest} \sim \IGamd\left(a_\kap + \frac{I\,K}{2}, b_\kap + \frac12\sum_{i}(\etav_i-\tev)^\trans(\etav_i-\tev) \right)\,.
	$$
	
	\item  Sample $\theta_j^{(b + 1)}$, $j = 1,\ldots,J$, according to an adaptive Metropolis-Hastings algorithm with the full conditional distribution:
	\begin{align*}
	p(\theta_j\mid\text{rest}) &\propto \prod_{i,i':i<i'} \Ber\left(\Phi(\zeta_j -  e^{\theta_j} \|\, \uv_{i,j} - \uv_{i',j} \,\|)\right) \times \Nor_1(\theta_j\mid \mu_\theta, \tau^2_\theta)\,.
	\end{align*}
	
	\item Sample $\mu_\theta^{(b+1)}$ from:
	$$
	\mu_\theta\mid\text{rest} \sim \Nor_1\left(\left[\frac{1}{v_\theta^2} + \frac{J}{\tau^2_\theta}\right]^{-1}\left[\frac{m_\theta}{v_\theta^2} + \frac{\sum_j \theta_j}{\tau^2_\theta}\right],\left[\frac{1}{v_\theta^2} + \frac{J}{\tau^2_\theta} \right]^{-1}\right)\,.
	$$
	
	\item Sample $(\tau^2_\theta)^{(b+1)}$ from:
	$$
	\tau^2_\theta\mid\text{rest} \sim \IGamd\left(a_\theta + \frac{J}{2}, b_\theta + \frac12\sum_j(\theta_j-\mu_\theta)^2\right)\,.
	$$
			
	\item  Sample $\zeta_j^{(b + 1)}$, $j = 1,\ldots,J$, according to an adaptive Metropolis-Hastings algorithm with the full conditional distribution:
	\begin{align*}
	p(\zeta_j\mid\text{rest}) &\propto \prod_{i,i':i<i'} \Ber\left(\Phi(\zeta_j -  e^{\theta_j} \|\, \uv_{i,j} - \uv_{i',j} \,\|)\right) \times \Nor_1(\zeta_j\mid \mu_\zeta, \tau^2_\zeta)\,.
	\end{align*}
	
	\item Sample $\mu_\zeta^{(b+1)}$ from:
	$$
	\mu_\zeta\mid\text{rest} \sim \Nor_1\left(\left[\frac{1}{v_\zeta^2} + \frac{J}{\tau^2_\zeta}\right]^{-1}\left[\frac{m_\zeta}{v_\zeta^2} + \frac{\sum_j \zeta_j}{\tau^2_\zeta}\right],\left[\frac{1}{v_\zeta^2} + \frac{J}{\tau^2_\zeta} \right]^{-1}\right)\,.
	$$
	
	\item Sample $(\tau^2_\zeta)^{(b+1)}$ from:
	$$
	\tau^2_\zeta\mid\text{rest} \sim \IGamd\left(a_\zeta + \frac{J}{2}, b_\zeta + \frac12\sum_j(\zeta_j-\mu_\zeta)^2\right)\,.
	$$
	
\end{enumerate}

\section{Notation}

The cardinality of a set $A$ is denoted by $|A|$. If P is a logical proposition, then $\ind{\text{P}} = 1$ if P is true, and $\ind{\text{P}} = 0$ if P is false. $\floor{x}$ denotes the floor of $x$, whereas $[n]$ denotes the set of all integers from 1 to $n$, i.e., $\{1,\ldots,n\}$. The Gamma function is given by $\Gamma(x)=\int_0^\infty u^{x-1}\,e^{-u}\,\text{d}u$.

Matrices and vectors with entries consisting of subscripted variables are denoted by a boldfaced version of the letter for that variable. For example, $\xv = (x_1,\ldots, x_n)$ denotes an $n\times1$ column vector with entries $x_1,\ldots, x_n$. We use $\zerov$ and $\onev$ to denote the column vector with all entries equal to 0 and 1, respectively, and $\I$ to denote the identity matrix. A subindex in this context refers to the corresponding dimension; for instance, $\I_n$ denotes the $n\times n$ identity matrix. The transpose of a vector $\xv$ is denoted by $\xv^\trans$; analogously for matrices. Moreover, if $\X$ is a square matrix, we use $\tr(\X)$ to denote its trace and $\X^{-1}$ to denote its inverse. The norm of $\xv$, given by $\sqrt{\xv^\trans\xv}$, is denoted by $\|\,\xv\|\,$.

\end{document}